
\magnification \magstep1
\raggedbottom
\openup 4\jot
\voffset6truemm
\headline={\ifnum\pageno=1\hfill\else
\hfill {\it Local Boundary Conditions for the Dirac Operator ...}
\hfill \fi}
\rightline {DAMTP R-90/26}
\vskip 1cm
\centerline {\bf LOCAL BOUNDARY CONDITIONS FOR THE DIRAC OPERATOR}
\centerline {\bf AND ONE-LOOP QUANTUM COSMOLOGY$^{*}$}
\vskip 1cm
\centerline {\bf Peter D. D'Eath$^{(a)}$ and Giampiero
Esposito$^{(a,b)}$}
\vskip 1cm
\centerline {$^{(a)}$ Department of Applied Mathematics and Theoretical
Physics}
\centerline {Silver Street, Cambridge CB3 9EW, U. K.}
\centerline {$^{(b)}$ St. John's College, Cambridge CB2 1TP, U. K.}
\vskip 1cm
\centerline {November 1990}
\vskip 1cm
\noindent
{\bf Abstract.} This paper studies local boundary conditions for fermionic
fields in quantum cosmology, originally introduced by
Breitenlohner, Freedman and Hawking for gauged supergravity theories in
anti-de Sitter space. For a spin-${1\over 2}$ field the conditions
involve the
normal to the boundary and the undifferentiated field.
A first-order differential operator for this Euclidean boundary-value
problem exists which is symmetric and has self-adjoint extensions.
The resulting eigenvalue equation in the case
of a flat Euclidean background with a three-sphere boundary of
radius $a$ is found to be :
$F(E)=[J_{n+1}(Ea)]^{2}-[J_{n+2}(Ea)]^{2}=0 , \forall
n \geq 0$. Using the theory of canonical products, this function $F$
may be expanded in terms of squared eigenvalues, in a way which has been used
in other recent one-loop calculations involving
eigenvalues of second-order operators. One can then study the
generalized Riemann $\zeta$-function formed from these squared eigenvalues.
The value of $\zeta(0)$ determines the scaling
of the one-loop prefactor in the Hartle-Hawking amplitude in quantum
cosmology. Suitable contour formulae, and the uniform
asymptotic expansions of the Bessel functions $J_{m}$ and their first
derivatives $J_{m}'$, yield for a massless Majorana field :
$\zeta(0)={11\over 360}$. Combining this with $\zeta(0)$ values
for other spins, one can then check whether the one-loop divergences
in quantum cosmology cancel in a supersymmetric theory.
\vskip 11cm
\noindent
$^{*}$Phys. Rev. D {\bf 43}, 3234 (1991).
\vskip 10cm
\centerline {\bf I. INTRODUCTION}
\vskip 1cm
In the last few years, a number of authors have investigated the one-loop
approximation in quantum cosmology and the boundary terms in the
asymptotic expansion of the heat kernel for fields of various spins.$^{1-11}$
For bosonic fields the most natural boundary conditions are local : either
Dirichlet or Neumann, or perhaps a combination of the two.$^{11,12}$
In the case of
fermionic fields one has a choice of local or non-local boundary
conditions.$^{8}$
The possibility of non-local boundary conditions arises because of the
first-order nature of the fermionic operators (a precise mathematical
treatment of the Dirac operator can be found in Ref. 13).
For example, take a
quantum cosmological model in which the Dirac field is
regarded as a perturbation around a Friedmann background gravitational model
containing a family of three-spheres of radius $a(t)$.$^{14}$
Using two-component spinor notation, the unprimed spin-${1\over 2}$ field
$\psi_{A}$ on a given three-sphere may be split into a sum
$\psi_{A}^{(+)}+\psi_{A}^{(-)}$, where $\psi_{A}^{(+)}$ is a sum of harmonics
having positive eigenvalues for the three-dimensional Dirac operator
${_{e}n_{AA'}}e^{BA'j}{ }^{(3)}D_{j}$
on the $S^3$, and $\psi_{A}^{(-)}$ is a sum
of harmonics having negative eigenvalues. Here ${_{e}n_{AA'}}$ is the spinor
version of the unit Euclidean normal$^{14}$
to the three-sphere, $e^{BA'j}$ is
the spinor version of the orthonormal spatial triad on the three-sphere,
and ${ }^{(3)}D_{j}$ is the three-dimensional covariant derivative
($j=1,2,3$).$^{14}$
A similar decomposition may be applied to the primed field
${\widetilde \psi}_{A'}$, which is taken to be independent of $\psi_{A}$, not
related by any conjugation operation. Boundary conditions suitable for
investigating the Hartle-Hawking quantum state$^{15,16}$
may be found by studying the classical version of the Hartle-Hawking path
integral, i. e. by asking for data on a three-sphere
bounding a compact region with a Riemannian metric, such that the classical
Dirac equation is well-posed. For a massless field, if one uses
spectral boundary conditions, one is forced to specify
$\psi_{A}^{(+)}$ and ${\widetilde \psi}_{A'}^{(+)}$ on the boundary, and
not $\psi_{A}^{(-)}$ and ${\widetilde \psi}_{A'}^{(-)}$. Heat-kernel
calculations relevant to the one-loop quantum amplitude with such non-local
boundary conditions are described in a subsequent paper.$^{17}$

Alternatively, one can examine possible local boundary
conditions for fermionic fields (particularly for spin ${1\over 2}$). For a
spin-${1\over 2}$ field $\Bigr(\psi_{A},\; {\widetilde \psi}_{A'}\Bigr)$ in
Riemannian space, referred to here as a Majorana spin-${1\over 2}$
field,$^{18}$ these conditions are
$$
\sqrt{2} \; {_{e}n_{A}^{\; \; A'}}\psi^{A}=\epsilon \;
{\widetilde \psi}^{A'} \; \; \; \; ,
\eqno (1.1)
$$
on the bounding surface. We use the word Majorana loosely : quantum
amplitudes computed formally in Lorentzian geometries, using Lorentzian
Majorana spinors $\Bigr(\psi_{A}, \; {\overline \psi}_{A'}\Bigr)$, where a
{\it bar} denotes complex conjugation, can be continued analytically to the
Euclidean regime by replacing ${\overline \psi}_{A'}$ by
${\widetilde \psi}_{A'}$, which is freed from being the conjugate of
$\psi_{A}$. Recently, a different definition of Euclidean Majorana spinors
has been discussed in Ref. 19; our use of the word Majorana is different
from that in Ref. 19.

In the boundary conditions (1.1) $_{e}n_{A}^{\; \; A'}$ is again the
Euclidean normal, and $\epsilon$ will be taken to be either $+1$ or $-1$.
Boundary conditions of the kind (1.1) have also
been studied in the present context in Ref. 12, where (though
using a different formalism) the more general possibility
$\epsilon =e^{i \theta}$ has been considered, with $\theta$
taken to be a real function of position on the boundary. However, the results
of Sec. II on self-adjointness for the Dirac problem only hold in the case
of real $\epsilon$, and attention will be restricted to this case.
The special case $\epsilon = \pm 1$ is part of a set
of boundary conditions introduced by Breitenlohner, Freedman
and Hawking$^{20,21}$
for gauged supergravity theories in anti-de Sitter
(hereafter referred to as ADS) space. The conditions (1.1) are generalized to
higher spins in an obvious way by requiring for spin $1$ that
$$
2 \; {_{e}n_{A}^{\; \; A'}}\; {_{e}n_{B}^{\; \; B'}}\psi^{AB}=\epsilon
\; {\widetilde \psi}^{A'B'} \; \; \; \; ,
\eqno (1.2)
$$
where $\Bigr(\psi_{AB},\; {\widetilde \psi}_{A'B'}\Bigr)$ is the
Maxwell field strength.
Similar boundary conditions may be written down for the
spin-${3\over 2}$ symmetric field strength
$\Bigr(\psi_{ABC}, \; {\widetilde \psi}_{A'B'C'}\Bigr)$
and for the Weyl spinor
$\Bigr(\psi_{ABCD}, \; {\widetilde \psi}_{A'B'C'D'}\Bigr)$ in the case
of spin $2$. For
a complex scalar field $\phi$, the conditions require the vanishing of
$Re \; \phi$ and ${\partial \over \partial n}Im \; \phi$, or of
$Im \phi$ and ${\partial \over \partial n}Re \; \phi$. As with the spectral
boundary conditions above, the classical massless
Dirac equation for $\Bigr(\psi_{A}, \; {\widetilde \psi}_{A'}\Bigr)$
is well-posed in a compact Riemannian region bounded by a
surface, on which the quantity $\Bigr(\sqrt{2} \;
{_{e}n_{A}^{\; \; A'}}\psi^{A}- \epsilon {\widetilde \psi}^{A'}
\Bigr)$ appearing in Eq. (1.1) is specified (at least in the case of a
spherically symmetric geometry). Similar results hold for other spins.

The relevance to gauged supergravity of the set of local boundary
conditions including (1.1,2) is as follows. ADS can be seen as the
maximally supersymmetric solution of the $O(N)$ gauged supergravity
theories. It has topology $S^{1}$x$R^{3}$, and its closed timelike
curves can be removed by considering its covering space CADS. CADS is
conformally flat, conformally embedded in the Einstein static universe,
and its boundary is the product of the time axis of the Einstein universe
and the two-sphere. The solutions of the twistor equation,$^{22}$
subject to a
suitable boundary condition,$^{21}$ generate the rigid supersymmetry
transformations between massless linearized fields of different spins
on an ADS background. In Ref. 20 the rigid supersymmetry
transformations map classical solutions of the linearized field
equations, subject to boundary conditions of the type (1.1,2) at infinity,
to classical solutions for an adjacent spin, again obeying the boundary
conditions at infinity. The set of boundary conditions including (1.1,2)
is thus in a certain sense supersymmetric.

Motivated by quantum cosmology, we examine fields on a flat
Euclidean background bounded by a three-sphere of radius $a$. There are
again$^{23}$ solutions of the twistor equation,
subject to a
certain boundary condition, which generate rigid supersymmetry
transformations among classical solutions obeying boundary conditions of
the type (1.1,2) on the bounding $S^3$. However,
these rigid transformations do not map {\it eigenfunctions} of the
spin-$s$ wave operators to eigenfunctions for adjacent spin
$s\pm {1\over 2}$ with the same eigenvalues.
Hence no cancellation can be expected {\it a priori}
between adjacent spins in a one-loop calculation of the
functional determinant about flat space (with $S^3$ boundary) for a
supersymmetric theory.

A different type of supersymmetric boundary condition is suggested by the
work of Ref. 24. In simple supergravity
the spatial tetrad $e_{\; \; \; \; \; i}^{AA'}$ (from which the intrinsic
spatial metric $h_{ij}$ is constructed) and the projection
$\Bigr(\epsilon {\widetilde \psi}_{i}^{A'}
-\sqrt{2} \; {_{e}n_{A}^{\; \; A'}}\psi_{i}^{A}\Bigr)$
formed from the spatial components
$\Bigr(\psi_{i}^{A}, \; {\widetilde \psi}_{i}^{A'}\Bigr)$ of the
spin-${3\over 2}$ potential, transform into each other under half
of the local
supersymmetry transformations at the boundary. Further, the supergravity
action, suitably modified by a boundary term, is invariant under this class
of local supersymmetry transformations. One is thus led to specify
$e_{\; \; \; \; \; i}^{AA'}$ and
$\Bigr(\epsilon {\widetilde \psi}_{i}^{A'}-
\sqrt{2} \; {_{e}n_{A}^{\; \; A'}}
\psi_{i}^{A}\Bigr)$ on the boundary in computing the quantum amplitude.
One could further check, along the lines of Ref. 25,
whether any local one-loop surface
counterterms are permitted by this remaining local supersymmetry,
i. e. whether there is any cancellation between the
one-loop determinants for spin $2$ and spin ${3\over 2}$.
Correspondingly, in studying the path integral, one arrives at the
consideration of these local
boundary conditions$^{11}$ by requiring that transition
amplitudes are invariant under BRST transformations (so that results do
not depend on the gauge-fixing term) and that supersymmetry is respected.
Extending
this to $O(N)$ supergravity, one finds, following the usual supersymmetry
transformation laws, that for spin-${1\over 2}$ fields
$\Bigr(\epsilon {\widetilde \psi}^{A'}-
\sqrt{2}\; {_{e}n_{A}^{\; \; A'}}\psi^{A}
\Bigr)$ should again be specified on the boundary, as in Eq. (1.1).
However, for spins higher than ${1\over 2}$
these boundary conditions are typically different from
those of Breitenlohner, Freedman and Hawking, involving projections of
potentials rather than field strengths.

Sec. II demonstrates self-adjointness
of the spin-${1\over 2}$ local boundary-value problem.
Sec. III examines the boundary
conditions (1.1) for a massless Majorana spin-${1\over 2}$ field on a flat
Euclidean background, bounded by a three-sphere of radius $a$.
The eigenvalue equation arising in the evaluation of the
one-loop functional determinant is derived. We then turn
to the generalized Riemann $\zeta$-function formed from the squared
eigenvalues. The value of $\zeta(0)$ measures the one-loop divergence
of the quantum amplitude prescribed by the given boundary conditions.
It also determines the scaling or $a$-dependence of the one-loop amplitude,
which is proportional to $a^{\zeta(0)}$ for a bosonic field and
$a^{-\zeta(0)}$ for a fermionic field (in the case of a scale-independent
measure). In Sec. IV the general structure of the $\zeta(0)$ calculation for
spin ${1\over 2}$ is described. This involves generalizing previous
work for bosonic fields,$^{4,8}$ and the calculation
depends on the theory of canonical products. The value of $\zeta(0)$ is
then computed in terms of its various contributions in Secs. V-VIII.
The result $\zeta(0)={11\over 360}$ for a massless Majorana spin-${1\over 2}$
field disagrees with the result ${17\over 180}$ of a
recent calculation,$^{11}$ which uses a quite different
approach. A discussion of this problem is given in Sec. IX, together
with other concluding remarks.
\vskip 1cm
\centerline {\bf II. SELF-ADJOINTNESS OF THE BOUNDARY-VALUE PROBLEM}
\vskip 1cm
We study the Hartle-Hawking path integral $^{14,15}$
$$
K_{HH} =\int e^{-I_{E}}D\psi^{A}
D{\widetilde \psi}^{A'}
\; \; \; \; ,
\eqno (2.1)
$$
taken over the class of spin-${1\over 2}$ fields $\Bigr(\psi^{A}, \;
{\widetilde \psi}^{A'}\Bigr)$ which obey Eq. (1.1) on the bounding $S^3$.
Here, with our conventions (see below),
$$
I_{E}={i\over 2}\int d^{4}x\sqrt{g}\left[\widetilde
\psi^{A'}\left(\nabla_{AA'}\psi^{A}\right)-\left(\nabla_{AA'}\widetilde
\psi^{A'}\right)\psi^{A}\right] \; \; \; \; ,
\eqno (2.2)
$$
is the Euclidean action for a massless spin-${1\over 2}$ field
$\Bigr(\psi^{A}, \; {\widetilde \psi}^{A'}\Bigr)$. The fields are defined
on the ball in Euclidean four-space bounded by a three-sphere of radius
$a$, subject to the local boundary conditions (1.1). The unprimed and primed
spinors are taken to transform under independent groups $SU(2)$ and
${\widetilde {SU(2)}}$, appropriate to Euclidean space. The fermionic fields
are taken to be anti-commuting, and Berezin integration is being used.$^{14}$

As with a bosonic one-loop path integral, one would like to be able to
express the path integral (2.1) in terms of a suitable product of eigenvalues.
The eigenvalue equations naturally arising from variation of the action
(2.2) are
$$
\nabla_{AA'}\psi_{n}^{A}=\lambda_{n} \widetilde \psi_{nA'} \; \; \; \; ,
\eqno (2.3)
$$
$$
\nabla_{AA'}\widetilde \psi_{n}^{A'}=\lambda_{n}\psi_{nA} \; \; \; \; ,
\eqno (2.4)
$$
subject again to the boundary conditions (1.1). One would expect
to expand a
typical pair $\Bigr(\psi^{A}, \; {\widetilde \psi}^{A'}\Bigr)$ in a complete
set of eigenfunctions (assuming provisionally that such a
complete set exists); one also
needs the cross-terms in $I_{E}$ to vanish. The gaussian fermionic path
integral (2.1) is then formally proportional to the
product of the eigenvalues
${\prod_{n}\left({\lambda_{n}\over {\widetilde \mu}}\right)}$, where
the constant ${\widetilde \mu}$ with dimensions of mass has been introduced
in order to make the product dimensionless.$^{26,27}$ In fact the
eigenvalues $\lambda_{n}$ for this problem are purely imaginary in the case
of a general Riemannian four-manifold with boundary. Further, in the
particular example of Euclidean four-space bounded by a three-sphere, the
eigenvalues occur in equal and opposite pairs $\pm \lambda_{n}$,
so that the formal product
${\prod_{n}\left({\lambda_{n}\over {\widetilde \mu}}\right)}$ can
instead be written as ${\prod_{n}\left({{\mid \lambda_{n} \mid}\over
{\widetilde \mu}}\right)}$, a product of positive real numbers. This formal
expression for the path integral (2.1) must then be regularized using
zeta-function methods.

The typical cross-term $\Sigma$ appearing in $I_{E}$ is
$$
\Sigma=
{i\over 2}\int d^{4}x \sqrt{g}\Bigr[(\lambda_{n}+\lambda_{m})
\widetilde \psi_{m}^{A'}\widetilde \psi_{nA'}
+(\lambda_{n}+\lambda_{m})\psi_{n}^{A}\psi_{mA}\Bigr] \; \; \; \; .
\eqno (2.5)
$$
Using the eigenvalue equations (2.3,4), the square bracket in Eq. (2.5) may
be rewritten as
$a\nabla_{AA'}\left(\psi_{m}^{A}
\widetilde \psi_{n}^{A'}\right)+b\nabla_{AA'}\left(\psi_{n}^{A}\widetilde
\psi_{m}^{A'}\right)$, with $a=-b={{\lambda_{n}+\lambda_{m}}
\over {\lambda_{n}-\lambda_{m}}}$, provided $\lambda_{n} \not = \lambda_{m}$.
Thus $\Sigma$ becomes
$$
\Sigma={i\over 2}
{(\lambda_{n}+\lambda_{m})\over (\lambda_{n}-\lambda_{m})}
\left[\int_{\partial M}d^{3}x \; \sqrt{h} \; {_{e}n_{AA'}}\psi_{m}^{A}
{\widetilde \psi}_{n}^{A'}
-\int_{\partial M}d^{3}x \; \sqrt{h} \; {_{e}n_{AA'}}\psi_{n}^{A}
{\widetilde \psi}_{m}^{A'}\right]=0 \; \; ,
\eqno (2.6)
$$
by virtue of the local boundary conditions (1.1), provided
$\lambda_{m} \not = \lambda_{n}$. In the degenerate case where the eigenvalues
are equal, a linear transformation within the degenerate eigenspace can be
found such that the cross-terms again vanish.

The property that $I_{E}$ can be written as a diagonal expression
in terms of a sum over eigenfunctions suggests that the Dirac action
used here, subject to local boundary conditions, can be expressed in terms
of a self-adjoint differential operator acting on fields
$\Bigr(\psi^{A}, \; {\widetilde \psi}^{A'}\Bigr)$. We shall see that this
is indeed the case, and
that the eigenvalues $\lambda_{n}$ are all purely imaginary.
The proof will be described in the case of a flat background geometry, but
can readily be generalized to the case of curved space.

Consider the space of spinor fields such as
$$
w\equiv \left(\psi^{A}, \; \widetilde \psi^{A'}\right)  \; \; \; \; ,
\; \; \; \;
z\equiv \left(\phi^{A}, \; \widetilde \phi^{A'}\right) \; \; \; \; ,
\eqno (2.7)
$$
defined on the ball of radius $a$ in Euclidean four-space, subject
to the boundary conditions (1.1) (and to suitable differentiability
conditions, to be specified later). Considering the action (2.2)
and eigenvalue equations (2.3,4), one is led to study the map
$$
C\; : \left(\psi^{A}, \; \widetilde \psi^{A'}\right)\rightarrow
\left(\nabla_{\; \; B'}^{A}\widetilde \psi^{B'},
\nabla_{B}^{\; \; A'}\psi^{B}\right) \; \; \; \; .
\eqno (2.8)
$$
Since the notion of self-adjointness involves the idea of reality, we also
need to introduce a conjugation operation on Euclidean spinors,$^{28}$ the
{\it dagger} operation
$$
{\left(\psi^{A}\right)}^{+}\equiv \epsilon^{AB}\delta_{BA'}
{\overline \psi}^{A'}  \; \; \; \; , \; \; \; \;
{\left({\widetilde \psi}^{A'}\right)}^{+} \equiv
\epsilon^{A'B'}\delta_{B'A}{\overline {\widetilde \psi}}^{A}
\; \; \; \; .
\eqno (2.9)
$$
Here $\delta_{BA'}$ is an identity matrix preserved by
$SU(2)$ transformations, and the alternating spinor $\epsilon^{AB}$
realizes the isomorphism
between spin space and its dual, raising and lowering indices according
to the rules : $\lambda^{A}=\epsilon^{AB}\lambda_{B}$, $\lambda_{A}=
\lambda^{B}\epsilon_{BA}$ (and similarly for $\epsilon^{A'B'}$).
Moreover, the {\it bar} symbol
$\overline {\psi^{A}}={\overline \psi}^{A'}$ denotes the usual complex
conjugation of $SL(2,C)$ spinors.$^{22}$ Note that the {\it dagger}
operation has the property
$$
{\left({\left(\psi^{A}\right)}^{+}\right)}^{+}=
\epsilon^{AC}\delta_{CB'}\overline {\left(\psi^{B+}\right)'}
=\epsilon^{AC}\delta_{CB'}\epsilon^{B'D'}\delta_{D'F}\psi^{F}
=-\psi^{A} \; \; \; \; ,
\eqno (2.10)
$$
and hence is anti-involutory.

{}From now on in this Section, despite the
requirement that spinors in the path integral be anti-commuting
Grassmann quantities, we study commuting spinors, for simplicity of
exposition of the self-adjointness. It may easily be checked that the
{\it dagger} operation has the following properties :
$$
(\psi_{A}+\lambda \phi_{A})^{+}=\psi_{A}^{+}+\lambda^{*}\phi_{A}^{+}
\; \; \; \; ,
\eqno (2.11)
$$
$$
\epsilon_{AB}^{+}=\epsilon_{AB} \; \; \; \; , \; \; \; \;
(\psi_{A}\phi_{B})^{+}=\psi_{A}^{+}\phi_{B}^{+}
\; \; \; \; ,
\eqno (2.12)
$$
$$
(\psi_{A})^{+}\psi^{A}>0 \; , \;
\forall \; \psi_{A}\not = 0 \; \; \; \; ,
\eqno (2.13)
$$
where the symbol $*$ denotes complex conjugation of scalars.
We can now define the scalar product
$$
(w,z)\equiv \int_{M}\left[\psi_{A}^{+}\phi^{A}+{\widetilde \psi_{A'}}^{+}
{\widetilde \phi}^{A'}\right]\sqrt{g}d^{4}x \; \; \; \; .
\eqno (2.14)
$$
This is indeed a scalar product, because it satisfies the following
properties, for all vectors $u$, $v$, $w$ and $\forall \lambda \in \; C$,
$$
(u,u)>0 \; \; \forall u \not =0 \; \; \; \; ,
\eqno (2.15)
$$
$$
(u,v+w)=(u,v)+(u,w) \; \; \; \; ,
\eqno (2.16)
$$
$$
(u,\lambda v)=\lambda(u,v) \; \; \; \; , \; \; \; \; \; \; \; \;
(\lambda u,v)={\lambda}^{*}(u,v) \; \; \; \; ,
\eqno (2.17)
$$
$$
(v,u)={(u,v)}^{*} \; \; \; \; .
\eqno (2.18)
$$
It will turn out that the operator $iC$ is symmetric using this scalar
product, i. e. that $(iCz,w)=(z,iCw)$ $\forall z,w$. This result will be
used in the course of proving further that the operator $iC$ has
self-adjoint extensions.

Let us now compute $(Cz,w)$ and $(z,Cw)$ for typical vectors $w$ and $z$
given by Eq. (2.7). From the definitions,
$$
(Cz,w)=
\int_{M}{\left(\nabla_{AB'}\phi^{A}\right)}^{+}{\widetilde \psi}^{B'}
\sqrt{g}d^{4}x
+\int_{M}{\left(\nabla_{BA'}{\widetilde \phi}^{A'}\right)}^{+}
\psi^{B}\sqrt{g}d^{4}x \; \; .
\eqno (2.19)
$$
Similarly, but integrating by parts, one finds
$$ \eqalignno{
(z,Cw)&=
\int_{M}\left(\nabla_{AB'}\phi^{A+}\right){\widetilde \psi}^{B'}\sqrt{g}
d^{4}x
+\int_{M}\left(\nabla_{BA'}\left({\widetilde \phi}^{A'}\right)^{+}\right)
\psi^{B}\sqrt{g}d^{4}x \cr
&-\int_{\partial M}(_{e}n_{AB'})\phi^{A+}{\widetilde \psi}^{B'}
\sqrt{h}d^{3}x
-\int_{\partial M}(_{e}n_{BA'}){\left({\widetilde \phi}^{A'}\right)}
^{+}\psi^{B}\sqrt{h}d^{3}x \; \; \; \; .
&(2.20)\cr}
$$
This may now be simplified using Eq. (2.9), the identity
$$
{\left(_{e}n^{AA'}\phi_{A}
\right)}^{+}=
\epsilon^{A'B'}\delta_{B'C}\;{\overline {_{e}n^{DC'}}}\;
{\overline {\phi_{D}}}=-\epsilon^{A'B'}\delta_{B'C}\left(_{e}n^{CD'}\right)
{\overline \phi}_{D'} \; \; \; \; ,
\eqno (2.21)
$$
and the boundary conditions on $S^{3}$ : $\sqrt{2}\; {_{e}n^{CB'}}\psi_{C}
={\widetilde \psi}^{B'}$, $\sqrt{2}\; {_{e}n^{AA'}}\phi_{A}
={\widetilde \phi}^{A'}$. The sum of the
boundary terms in Eq. (2.20) vanishes. Therefore, in verifying that
$(iCz,w)=(z,iCw)$, it is sufficient to check that the corresponding volume
integrands are equal. This involves detailed use of the Infeld-van der
Waerden connection symbols $\sigma_{\; \; AA'}^{a}$ and
$\sigma_{a}^{\; \; AA'}$, where in particular we take
$\sigma_{0}=-{i\over \sqrt{2}}I$,
$\sigma_{i}={\Sigma_{i} \over \sqrt{2}}$ ($i=1,2,3$), where $\Sigma_{i}$ are
the Pauli matrices.
In simplifying Eqs. (2.19,20) one uses, for example, the flat-space relations
${\left(\nabla_{BA'}{\widetilde \phi}^{A'}\right)}^{+}=\delta_{BF'}
{\overline \sigma}_{\; \; C}^{F'\; \; a}\partial_{a}
\left({\overline {\widetilde \phi}}^{C}\right)$ and
$\left(\nabla_{BA'}\left({\widetilde \phi}^{A'}\right)^{+}\right)=
-\delta_{CF'}\sigma_{B}^{\; \; F'a}\partial_{a}
\left({\overline {\widetilde \phi}}^{C}\right)$. A straightforward but
tedious calculation leads to the desired relation
$(iCz,w)=(z,iCw)$, $\forall w,z$.

Every symmetric
operator has a closure, and the operator and its closure have the same closed
extensions. Moreover, a closed symmetric operator on a
Hilbert space is self-adjoint if and only if its spectrum is a subset of the
real axis (see Ref. 29, p 136).
To prove self-adjointness for our boundary-value problem, we recall a result
due to Von Neumann (see Ref. 29, p 143) : given a symmetric operator $A$
with domain $D(A)$, and a map $F : D(A)\rightarrow D(A)$ such that
$$
F(\alpha w + \beta z)=\alpha^{*}F(w)+\beta^{*}F(z) \; \; \; \; ,
\eqno (2.22)
$$
$$
(w,w)=(Fw,Fw) \; \; \; \; ,
\eqno (2.23)
$$
$$
F^{2}=\pm I \; \; \; \; ,
\eqno (2.24)
$$
$$
FA=AF \; \; \; \; ,
\eqno (2.25)
$$
then $A$ has self-adjoint extensions. In the present case (see Eq. (2.9))
let $D$ denote the operator
$$
D : \left(\psi^{A}, \; {\widetilde \psi}^{A'}\right)\rightarrow
\left(\psi^{A+}, {\left({\widetilde \psi}^{A'}\right)}^{+}\right)
\; \; \; \; ,
\eqno (2.26)
$$
and let $F=iD$ and $A=iC$. It can be checked that the operator $F$ maps
$D(A)$ to $D(A)$. Denoting by $k$ an integer $\geq 2$ and defining
$$
D(A)\equiv \left \{ \Bigr(\psi^{A}, \; {\widetilde \psi}^{A'}\Bigr) :
\psi^{A} \; and \; {\widetilde \psi}^{A'} \; are \; C^{k}, \; and \;
\sqrt{2} \; {_{e}n^{AA'}}\psi_{A}
=\epsilon \; {\widetilde \psi}^{A'} \; on \; S^{3} \right \} \; \; \; \; ,
\eqno (2.27)
$$
one finds that $F$ maps $\Bigr(\psi^{A}, \; {\widetilde \psi}^{A'}\Bigr)$ to
$\Bigr(\beta^{A}, \; {\widetilde \beta}^{A'}\Bigr)=
\left(i\Bigr(\psi^{A}\Bigr)^{+},i \Bigr({\widetilde \psi}^{A'}\Bigr)^{+}
\right)$ such that :
$$
\sqrt{2} \; {_{e}n^{AA'}}\beta_{A}=\gamma \; {\widetilde \beta}^{A'}
\; on \; S^{3} \; \; \; \; ,
\eqno (2.28)
$$
where $\gamma =\epsilon^{*}$. The boundary condition (2.28) is clearly
of the type which occurs in Eq. (2.27) provided $\epsilon$ is real,
and the differentiability of
$\Bigr(\beta^{A}, \; {\widetilde \beta}^{A'}\Bigr)$ is not affected by the
action of $F$. In deriving Eq. (2.28), we have used Eq. (2.21).
The requirement of self-adjointness enforces
the choice of a real function $\epsilon$, which for simplicity
we have taken in Sec. I to be a constant.
Moreover, in view of Eq. (2.10), Eqs. (2.22,24) hold when $F=iD$,
provided we write Eq. (2.24) as $F^{2}=-I$. Since, in Ref. 29, condition
(2.24) is written as $F^{2}=I$, and examples are later given
(see p 144 therein) where $F$ is complex conjugation, some explanation is in
order. Our problem is in the
Euclidean regime, where the only possible conjugation is
the {\it dagger} operation, which is anti-involutory on spinors with an odd
number of indices. Thus we are using a slight generalization of Von Neumann's
theorem. Note here that if $F : D(A)\rightarrow D(A)$ satisfies Eqs.
(2.22)-(2.25), then the same is clearly true of ${\widetilde F}=-iD=-F$.
Hence
$$
-F \; D(A) \subseteq D(A) \; \; \; \; ,
\eqno (2.29)
$$
$$
F \; D(A) \subseteq D(A)  \; \; \; \; .
\eqno (2.30)
$$
Acting with $F$ on both sides of Eq. (2.29), we find
$$
D(A) \subseteq F \; D(A) \; \; \; \; ,
\eqno (2.31)
$$
using the property $F^{2}=-I$. Eqs. (2.30,31) imply
that $F \; D(A)=D(A)$, so that $F$ takes $D(A)$ onto $D(A)$ in our case.
Comparison with the proof on p 144 of Ref. 29
shows that this is sufficient to generalize Von Neumann's
theorem to the Dirac operator with the boundary conditions (1.1).

It remains to verify conditions (2.23,25). First, note that
$$ \eqalignno{
\left(Fw,Fw\right)&=
\left(iDw,iDw\right) \cr
&=\int_{M}{\left(i\psi_{A}^{+}\right)}^{+}i \psi^{A+}\sqrt{g}d^{4}x
+\int_{M}{\left(i{\widetilde \psi}_{A'}^{+}\right)}^{+}i
{\left({\widetilde \psi}^{A'}\right)^{+}}\sqrt{g}d^{4}x\cr
&=\left(w,w\right) \; \; \; \; ,
&(2.32)\cr}
$$
using Eqs. (2.10,14), for commuting spinors. Second,
$$
FAw=
\left(iD\right)\left(iC\right)w
=i{\left[i\left(\nabla_{\; \; B'}^{A}{\widetilde \psi}^{B'},
\nabla_{B}^{\; \; A'}\psi^{B}\right)\right]}^{+}
={\left(\nabla_{\; \; B'}^{A}{\widetilde \psi}^{B'},
\nabla_{B}^{\; \; A'}\psi^{B}\right)}^{+} \; \; ,
\eqno (2.33)
$$
$$
AFw=
\left(iC\right)\left(iD\right)w
=iCi\left(\psi^{A+},{\left({\widetilde \psi}^{A'}\right)}^{+}\right)
=-\left(\nabla_{\; \; B'}^{A}{\left({\widetilde \psi}^{B'}\right)}^{+},
\nabla_{B}^{\; \; A'}\psi^{B+}\right) \; \; ,
\eqno (2.34)
$$
yielding Eq. (2.25), with the help of Eq. (2.9) and of the fact that the
complex conjugate of $\sigma_{\; \; A'}^{A \; \; \; a}$ is equal to
$\sigma_{A}^{\; \; A'a}$, $\forall a=0,1,2,3$.
To sum up, it has been shown that the operator
$iC$ arising in this boundary-value problem is symmetric and has self-adjoint
extensions. In particular, the eigenvalues $\lambda_{n}$ of $C$
are purely imaginary. The same result follows for a curved four-dimensional
Riemannian space with boundary, by a straightforward generalization.

Throughout this Section, we have considered only the first-order operator
$iC \; : \; \Bigr(\psi^{A}, \; {\widetilde \psi}^{A'}\Bigr)
\rightarrow \Bigr(i \nabla_{\; \; B'}^{A}{\widetilde \psi}^{B'}, \;
i\nabla_{B}^{\; \; A'}\psi^{B}\Bigr)$ which appears naturally in varying the
action (2.2), subject to the local boundary conditions (1.1), and whose
eigenvalues $i\lambda_{n}$ appear in the formal product expression
${\prod_{n}\left({\mid \lambda_{n} \mid \over {\widetilde \mu}}\right)}$
for the
path integral (2.1). An alternative procedure is, of course, to square up the
Dirac operator and study the second-order operator $C^{+}C$, which in our
flat background is just minus the Laplacian acting on spinors. This approach
has been taken, for example, in Refs. 9, 11, 12.
The path integral (2.1) can instead be evaluated in terms of eigenvalues
of $C^{+}C$, but an additional boundary condition, which can be written in
the form
$$
\sqrt{2} \; {_{e}n_{A}^{\; \; A'}}\nabla_{\; \; B'}^{A}
{\widetilde \psi}^{B'} = \epsilon \; \nabla_{B}^{\; \; A'}\psi^{B}
\; \; \; \; ,
\eqno (2.35)
$$
must be imposed together with Eq. (1.1). [In Refs. 9, 11, this condition
is written as
$$
\left({_{e}n^{BB'}}\nabla_{BB'}+{1\over 2}trK \right)
\Bigr(\sqrt{2}\;{_{e}n_{A}^{\; \; A'}}\psi^{A}+\epsilon {\widetilde \psi}^{A'}
\Bigr)=0 \; \; \; \; ,
$$
where $K_{ij}$ is the second fundamental form of the boundary].
This extra condition is automatically obeyed by the eigenfunctions
$\Bigr({\psi}_{n}^{A}, \; {\widetilde \psi}_{n}^{A'}\Bigr)$ of the
first-order problem. The eigenfunctions of the first- and second-order
problems coincide, and the eigenvalues of the second-order problem are of
course ${\mid \lambda_{n} \mid}^{2}$. The formal product expressions for
the path integral (2.1) agree. The extra condition (2.35) involves not just
$\psi^{A}$ and ${\widetilde \psi}^{A'}$ on the bounding surface, but also
their normal derivatives. When viewed in terms of the path integral subject
to the boundary conditions (1.1), condition (2.35) seems extraneous. We
have therefore preferred to use only the boundary conditions (1.1) and to
study the associated first-order Dirac operator.

The product is regularized by defining the zeta-function
$$
\zeta(s)\equiv \sum_{n,k}d_{k}(n) {\mid \lambda_{n,k} \mid}^{-2s}
\; \; \; \; .
\eqno (2.36)
$$
Here we modify the notation in anticipation of Secs. III, IV : the
eigenvalues $\lambda_{n,k}$ in the example studied there are labelled by two
integers $n$ and $k$, and the degeneracy $d_{k}(n)$ depends only on $n$.
Because the ${\mid \lambda_{n,k} \mid}^{2}$ are the eigenvalues for a
second-order self-adjoint problem, $\zeta(s)$ converges for $Re(s)>2$ and
can be analytically continued to a meromorphic function regular at the
origin, with poles only at $s={1\over 2},1,{3\over 2},2$ (see Eqs. (4.9,11)
of Ref. 30).
The formal expression $\log
{\prod_{n,k}\left({\mid \lambda_{n,k} \mid \over {\widetilde \mu}}
\right)}$ is then evaluated, following the standard procedure,$^{27,30}$ as
$-{1\over 2}\zeta'(0)-\zeta(0) \log {\widetilde \mu}$.
\vskip 10cm
\centerline {\bf III. LOCAL BOUNDARY CONDITIONS AND EIGENVALUE EQUATION}
\vskip 1cm
In this Section we consider the eigenvalue equation for a massless
Majorana field subject to the boundary conditions (1.1) on a three-sphere
of radius $a$, bounding a region of Euclidean four-space centred on the
origin. The field $\Bigr(\psi^{A}, \; {\widetilde \psi}^{A'}\Bigr)$ may be
expanded in terms of harmonics on the family of spheres centred on the
origin,$^{14}$ as
$$
\psi^A={\tau^{-{\textstyle {3 \over 2}}}\over 2\pi}\sum_{npq}
\alpha_{n}^{pq}\Bigr(m_{np}(\tau)\rho^{nqA} +{\widetilde r}_{np}(\tau)
{\overline \sigma}^{nqA}\Bigr) \; \; \; \; ,
\eqno (3.1)
$$
$$
{\widetilde \psi}^{A'}={\tau^{-{\textstyle {3 \over 2}}}\over 2\pi}
\sum_{npq} \alpha_{n}^{pq}\left({\widetilde m}_{np}(\tau)
{\overline \rho}^{nqA'}+
r_{np}(\tau) \sigma^{nqA'}\right) \; \; \; \; .
\eqno (3.2)
$$
Here $\tau$ is the radius of a three-sphere. In the summation, $n$ runs
from $0$ to $\infty$, $p$ and $q$ from $1$ to $(n+1)(n+2)$. The
$\alpha_{n}^{pq}$ are a collection of matrices introduced for
convenience, where, for each $n$, $\alpha_{n}^{pq}$ is block-diagonal
in the indices $pq$, with blocks
$\pmatrix {1&1\cr 1&-1\cr}$. The harmonics $\rho^{nqA}$ and
$\sigma^{nqA'}$ have positive eigenvalues ${1\over 2}(n+{3\over 2})$ of
the intrinsic three-dimensional Dirac operator on $S^3$, while the harmonics
${\overline \sigma}^{nqA}$ and ${\overline \rho}^{nqA'}$ have negative
eigenvalues $-{1\over 2}(n+{3\over 2})$. The {\it tilde} symbol
does not denote any conjugation operation : pairs such as
$m_{np},{\widetilde m}_{np}$ are independent functions of $\tau$.
The harmonics ${\overline \rho}^{nqA'}$ and
${\overline \sigma}^{nqA}$ may be re-expressed in terms of the harmonics
$n_{A}^{\; \; A'}\rho^{npA}$ and $n_{\; \; A'}^{A}\sigma^{npA'}$,
where $n^{AA'}=i \; _{e}n^{AA'}$ is the Lorentzian normal,$^{14}$
using the relations  $^{14}$
$$
\bar \rho^{nqA'}=2n_{A}^{\ A'}\sum_{d}\rho^{ndA} (A_{n}^{-1} H_{n})^{dq}
\; \; \; \; ,
\eqno (3.3)
$$
$$
{\overline \sigma}^{nqA}=2n_{\ A'}^{A}\sum_{d}\sigma^{ndA'}(A_{n}^{-1} H_{n})
^{dq} \; \; \; \; .
\eqno (3.4)
$$
Here the matrix $A_{n}^{-1}H_{n}$ of Ref. 14 is again block-diagonal in the
indices $dq$, for each $n$, with blocks
${1\over \sqrt{2}} \pmatrix {0&-1\cr 1&0\cr}$.

For simplicity, consider first that part of the boundary condition (1.1)
which involves the $\rho$ harmonics.
The relations (3.1)-(3.3) yield for each $n$ :
$$
-i\sum_{pq} \pmatrix {1&1\cr 1&-1\cr}^{pq}m_{np}(a)\rho^{nqA}=
\epsilon \sum_{pq} \pmatrix {1&1\cr 1&-1\cr}^{pq}\widetilde m_{np}(a)
\sum_{d}\rho^{ndA} \pmatrix {0&-1\cr 1&0\cr}^{dq}.
\eqno (3.5)
$$
In the typical case of the indices $p,q,d=1,2$, this implies
$$ \eqalignno{
-im_{n1}(a)\Bigr(\rho^{n1A}+\rho^{n2A}\Bigr)-im_{n2}(a)
\Bigr(\rho^{n1A}-\rho^{n2A}\Bigr)&=
\epsilon \widetilde m_{n1}(a)\Bigr(\rho^{n2A}-\rho^{n1A}\Bigr)\cr
&+\epsilon \widetilde m_{n2}(a)\Bigr(\rho^{n2A}+\rho^{n1A}\Bigr).&(3.6)\cr}
$$
Similar equations hold for adjacent indices $p,q,d=2k+1,2k+2$. Since the
harmonics $\rho^{n1A}$ and $\rho^{n2A}$ on the bounding three-sphere are
linearly independent, we have
$$
-i \Bigr[m_{n1}(a)+m_{n2}(a)\Bigr]=\epsilon \Bigr[\widetilde m_{n2}(a)
-{\widetilde m}_{n1}(a)\Bigr] \; \; \; \; ,
\eqno (3.7)
$$
$$
-i \Bigr[m_{n1}(a)-m_{n2}(a)\Bigr]=\epsilon \Bigr[\widetilde m_{n2}(a)
+{\widetilde m}_{n1}(a)\Bigr] \; \; \; \; ,
\eqno (3.8)
$$
whose solution is
$$
-im_{n1}(a)=\epsilon \widetilde m_{n2}(a) \; \; \; \; ,
\eqno (3.9)
$$
$$
im_{n2}(a)=\epsilon \widetilde m_{n1}(a) \; \; \; \; .
\eqno (3.10)
$$
In the same way, the part of Eq. (1.1) involving the $\sigma$ harmonics
leads to
$$
\epsilon
\sum_{pq} \pmatrix {1&1\cr 1&-1\cr}^{pq}r_{np}(a)\sigma^{nqA'}=
i\sum_{pq}\pmatrix {1&1\cr 1&-1\cr}^{pq}\widetilde r_{np}(a)
\sum_{d} \sigma^{ndA'}\pmatrix {0&-1\cr 1&0\cr}^{dq},
\eqno (3.11)
$$
which implies, for example
$$ \eqalignno{
\epsilon
\Bigr[r_{n1}(a)+r_{n2}(a)\Bigr]\sigma^{n1A'}+
\epsilon
\Bigr[r_{n1}(a)-r_{n2}(a)\Bigr]\sigma^{n2A'}&=
-i\Bigr[\widetilde r_{n1}(a)-\widetilde r_{n2}(a)\Bigr]\sigma^{n1A'}\cr
&+i\Bigr[\widetilde r_{n1}(a)+\widetilde r_{n2}(a)\Bigr]\sigma^{n2A'},
&(3.12)\cr}
$$
leading to
$$
-i \widetilde r_{n1}(a)=\epsilon r_{n2}(a) \; \; \; \; ,
\eqno (3.13)
$$
$$
i \widetilde r_{n2}(a)=\epsilon r_{n1}(a) \; \; \; \; ,
\eqno (3.14)
$$
and similar equations for other adjacent indices $p,q,d$. Thus, defining
$$
x \equiv m_{n1}, X \equiv m_{n2},
\widetilde x \equiv \widetilde m_{n1},
\widetilde X \equiv \widetilde m_{n2},
y \equiv r_{n1}, Y \equiv r_{n2},
\widetilde y \equiv \widetilde r_{n1},
\widetilde Y \equiv \widetilde r_{n2},
\eqno (3.15)
$$
we may cast Eqs. (3.9,10,13,14) in the form
$$
-ix(a)=\epsilon \widetilde X(a) \; \; \; \; , \; \; \; \; \; \; \; \; \;
iX(a)=\epsilon \widetilde x(a) \; \; \; \; ,
\eqno (3.16)
$$
$$
-i \widetilde y(a)=\epsilon Y(a) \; \; \; \; , \; \; \; \; \; \; \; \; \;
i \widetilde Y(a)=\epsilon y(a) \; \; \; \; .
\eqno (3.17)
$$
Again, similar equations hold relating $m_{np},{\widetilde m}_{np},
r_{np}$ and ${\widetilde r}_{np}$ at the boundary for adjacent indices
$p=2k+1,2k+2$ ($k=0,1,...,{n\over 2}(n+3)$).

Now we turn to the eigenvalue equations (2.3,4), described with the help
of the preceding decomposition. Studying the case of coupled modes with
adjacent indices $p,q=2k+1,2k+2$ for a given value of $n$, as above, we
write $l=n+{3\over 2}$ and $E=i \lambda_{n}=-Im(\lambda_{n})$. Introducing
$\forall n \geq 0$ the operators
$$
L_{n}\equiv {d\over d\tau}-{l\over \tau} \; \; \; \; , \; \; \; \;
M_{n}\equiv {d\over d\tau}+{l\over \tau} \; \; \; \; ,
\eqno (3.18)
$$
the eigenvalue equations are found to be
$$
L_{n}x=E{\widetilde x} \; \; \; \; , \; \; \; \;
M_{n}{\widetilde x}=-Ex \; \; \; \; ,
\eqno (3.19)
$$
$$
L_{n}y=E{\widetilde y} \; \; \; \; , \; \; \; \;
M_{n}{\widetilde y}=-Ey \; \; \; \; ,
\eqno (3.20)
$$
$$
L_{n}X=E{\widetilde X} \; \; \; \; , \; \; \; \;
M_{n}{\widetilde X}=-EX \; \; \; \; ,
\eqno (3.21)
$$
$$
L_{n}Y=E{\widetilde Y} \; \; \; \; , \; \; \; \;
M_{n}{\widetilde Y}=-EY \; \; \; \; .
\eqno (3.22)
$$
We now define $\forall n \geq 0$ the differential operators
$$
P_{n}\equiv {d^{2}\over d\tau^{2}}+
\left[E^{2}-{((n+2)^{2}-{1\over 4})\over \tau^{2}}\right]
\; \; \; \; ,
\eqno (3.23)
$$
$$
Q_{n}\equiv {d^{2}\over d\tau^{2}}+
\left[E^{2}-{((n+1)^{2}-{1\over 4})\over \tau^{2}}\right]
\; \; \; \; .
\eqno (3.24)
$$
Eqs. (3.19)-(3.22) lead straightforwardly to the following second-order
equations :
$$
P_{n}{\widetilde x}=P_{n}{\widetilde X}=P_{n}{\widetilde y}
=P_{n}{\widetilde Y}=0
\; \; \; \; ,
\eqno (3.25)
$$
$$
Q_{n}y=Q_{n}Y=Q_{n}x=Q_{n}X=0
\; \; \; \; .
\eqno (3.26)
$$
The solutions of Eqs. (3.25,26) which are regular at the origin are
$$
\widetilde x=C_{1}\sqrt{\tau}J_{n+2}(E\tau) \; \; \; \; , \; \; \; \; \; \;
\; \;
\widetilde X=C_{2}\sqrt{\tau}J_{n+2}(E\tau) \; \; \; \; ,
\eqno (3.27)
$$
$$
x=C_{3}\sqrt{\tau}J_{n+1}(E\tau) \; \; \; \; , \; \; \; \; \; \; \; \;
X=C_{4}\sqrt{\tau}J_{n+1}(E\tau) \; \; \; \; ,
\eqno (3.28)
$$
$$
\widetilde y=C_{5}\sqrt{\tau}J_{n+2}(E\tau) \; \; \; \; ,
\; \; \; \; \; \; \; \;
\widetilde Y=C_{6}\sqrt{\tau}J_{n+2}(E\tau) \; \; \; \; ,
\eqno (3.29)
$$
$$
y=C_{7}\sqrt{\tau}J_{n+1}(E\tau) \; \; \; \; , \; \; \; \; \; \; \; \;
Y=C_{8}\sqrt{\tau}J_{n+1}(E\tau) \; \; \; \; .
\eqno (3.30)
$$

In order to find the equation obeyed by $E$, we must now insert
Eqs. (3.27)-(3.30)
into the boundary conditions (3.16,17), taking into account also the
first-order system given by Eqs. (3.19)-(3.22). This gives the
following eight equations :
$$
-iC_{3}J_{n+1}(Ea)=\epsilon C_{2}J_{n+2}(Ea) \; \; \; \; ,
\eqno (3.31)
$$
$$
iC_{4}J_{n+1}(Ea)=\epsilon C_{1}J_{n+2}(Ea) \; \; \; \; ,
\eqno (3.32)
$$
$$
-iC_{5}J_{n+2}(Ea)=\epsilon C_{8}J_{n+1}(Ea)  \; \; \; \; ,
\eqno (3.33)
$$
$$
iC_{6}J_{n+2}(Ea)=\epsilon C_{7}J_{n+1}(Ea) \; \; \; \; ,
\eqno (3.34)
$$
$$
C_{1}=-{{EC_{3}J_{n+1}(Ea)}\over
{E\dot J_{n+2}(Ea)+(n+2){\textstyle {J_{n+2}(Ea)\over a}}}} \; \; \; \; ,
\eqno (3.35)
$$
$$
C_{2}=-{{EC_{4}J_{n+1}(Ea)}\over
{E\dot J_{n+2}(Ea)+(n+2){\textstyle {J_{n+2}(Ea)\over a}}}} \; \; \; \; ,
\eqno (3.36)
$$
$$
C_{7}={{EC_{5}J_{n+2}(Ea)}\over
{E\dot J_{n+1}(Ea)-(n+1){\textstyle {J_{n+1}(Ea)\over a}}}} \; \; \; \; ,
\eqno (3.37)
$$
$$
C_{8}={{EC_{6}J_{n+2}(Ea)}\over
{E\dot J_{n+1}(Ea)-(n+1){\textstyle {J_{n+1}(Ea)\over a}}}} \; \; \; \; .
\eqno (3.38)
$$
Note that these give separate relations among the constants $C_{1},C_{2},
C_{3},C_{4}$ and among $C_{5},C_{6},C_{7},C_{8}$. For example, eliminating
$C_{1},C_{2},C_{3},C_{4}$, using Eqs. (3.31,32,35,36)
and the useful identities $^{31}$
$$
Ea\dot J_{n+1}(Ea)-(n+1)J_{n+1}(Ea)=-EaJ_{n+2}(Ea) \; \; \; \; ,
\eqno (3.39)
$$
$$
Ea\dot J_{n+2}(Ea)+(n+2)J_{n+2}(Ea)=EaJ_{n+1}(Ea) \; \; \; \; ,
\eqno (3.40)
$$
one finds
$$
-i\epsilon {J_{n+1}(Ea)\over J_{n+2}(Ea)}={\epsilon}^{2}{C_{2}\over C_{3}}
={\epsilon}^{2}{C_{4}\over C_{1}}
=-i{\epsilon}^{3}{J_{n+2}(Ea)\over J_{n+1}(Ea)}
\; \; \; \; ,
\eqno (3.41)
$$
which implies (since $\epsilon=\pm 1$) :
$$
J_{n+1}(Ea)=\pm J_{n+2}(Ea) \; \forall n \geq 0 \; \; \; \; ,
\eqno (3.42)
$$
which is the desired set of eigenvalue conditions. Exactly the same set of
eigenvalue conditions would have arisen from eliminating
$C_{5},C_{6},C_{7},C_{8}$.

Note that, since ${J_{n+2}(z)\over J_{n+1}(z)}$ is an odd function of $z$,
eigenvalues occur in equal and opposite pairs $\pm E$. The degeneracy of a
particular eigenvalue $E$ corresponding to a given $n$ is $(n+1)(n+2)$, since
this is the number of indices $p$ in Eqs. (3.1,2) for a given $n$.

The limiting behaviour of
the eigenvalues can be found under certain approximations. For example, if
$n$ is fixed and $\mid z \mid \rightarrow \infty$, one has the standard
asymptotic expansion : $^{32}$
$$
J_{n}(z)\sim \sqrt{{2\over {\pi z}}}
\cos \left(z-{n\pi \over 2}-{\pi \over 4}\right)
+O{\Bigr(z^{-{3\over 2}}\Bigr)} \; \; \; \; .
\eqno (3.43)
$$
Thus, writing Eq. (3.42) in the form $J_{n+1}(E)=\kappa J_{n+2}(E)$, where
$\kappa = \pm 1$ and we set $a=1$ for simplicity, two asymptotic sets of
eigenvalues result :
$$
E^{+} \sim \pi \left({n\over 2}+L \right) \; \; if \; \; \kappa=1
\; \; \; \; ,
\eqno (3.44)
$$
$$
E^{-} \sim \pi \left ({n\over 2}+M+{1\over 2}\right) \; \; if \; \;
\kappa=-1 \; \; \; \; ,
\eqno (3.45)
$$
where $L$ and $M$ are large integers (both positive and negative).
One can also obtain an estimate of the smallest eigenvalues, for a given
large $n$. The asymptotic expansions in Sec. 9.3 of Ref. 32 show that these
eigenvalues have the asymptotic form
$$
\mid E \mid \sim \Bigr[n+o(n)\Bigr] \; \; \; \; .
\eqno (3.46)
$$

In general it is very difficult to solve Eq. (3.42) numerically,
because the recurrence relations which enable one to compute Bessel functions
starting from $J_{0}$ and $J_{1}$ are a source of large errors when the
argument is comparable with the order. However, a remarkable numerical
study has been carried out in Ref. 33. In that paper, the authors study
eigenvalues of the Dirac operator with local boundary conditions, in the
case of neutrino billiards. This corresponds to massless spin-${1\over 2}$
particles moving under the action of a potential describing
a hard wall bounding a finite domain. The authors end up with an eigenvalue
equation of the kind $J_{l}(k_{nl})=J_{l+1}(k_{nl})$, and compute the
lowest $2600$ positive eigenvalues $k_{nl}$.
\vskip 100cm
\centerline {\bf IV. GENERAL STRUCTURE OF THE $\zeta(0)$ CALCULATION FOR THE}
\centerline {\bf SPIN-${1\over 2}$ FIELD SUBJECT TO LOCAL BOUNDARY
CONDITIONS ON $S^3$}
\vskip 0.3cm
In Sec. III we derived the eigenvalue equation (3.42) which, setting
$a=1$ for simplicity, can be written in the non-linear form
$$
F(E)=[J_{n+1}(E)]^{2}-[J_{n+2}(E)]^{2}=0 \; \; \; \; \forall n \geq 0
\; \; \; \; .
\eqno (4.1)
$$
The function $F$ is the product of the entire functions
(functions analytic in the whole complex plane)
$F_{1}=J_{n+1}-J_{n+2}$ and $F_{2}=J_{n+1}+J_{n+2}$,
which can be written in the form
$$
F_{1}(z)=J_{n+1}(z)-J_{n+2}(z)=\gamma_{1}z^{(n+1)}e^{g_{1}(z)}
\prod_{i=1}^{\infty}\left(1-{z\over \mu_{i}}\right)e^{z\over \mu_{i}}
\; \; \; \; ,
\eqno (4.2)
$$
$$
F_{2}(z)=J_{n+1}(z)+J_{n+2}(z)=\gamma_{2}z^{(n+1)}e^{g_{2}(z)}
\prod_{i=1}^{\infty}\left(1-{z\over \nu_{i}}\right)e^{z\over \nu_{i}}
\; \; \; \; .
\eqno (4.3)
$$
In Eqs. (4.2,3), $\gamma_{1}$ and $\gamma_{2}$ are constants,
$g_{1}$ and $g_{2}$ are entire functions, the
$\mu_{i}$ are the (real) zeros of $F_{1}$ and the $\nu_{i}$ are the (real)
zeros of $F_{2}$. In fact, using the terminology in Ref. 34
(see pp 194-195 therein), $F_{1}$ and $F_{2}$ are entire functions whose
canonical product has genus 1. Namely, in light of the asymptotic behaviour
of the eigenvalues (see Eqs. (3.44,45)), we know that
$\sum_{i=1}^{\infty}{1\over \mid \mu_{i} \mid}=\infty$ and
$\sum_{i=1}^{\infty}{1\over \mid \nu_{i} \mid}=\infty$, whereas
$\sum_{i=1}^{\infty}{1\over {\mid \mu_{i} \mid}^{2}}$ and
$\sum_{i=1}^{\infty}{1\over {\mid \nu_{i} \mid}^{2}}$ are convergent. This is
why $e^{z\over \mu_{i}}$ and $e^{z\over \nu_{i}}$ must appear in Eqs.
(4.2,3), which are called the canonical-product representations of
$F_{1}$ and $F_{2}$. The genus of the canonical product for $F_{1}$ is the
minimum integer $h$ such that $\sum_{i=1}^{\infty}{1\over
{\mid \mu_{i} \mid}^{h+1}}$ converges, and similarly for $F_{2}$,
replacing $\mu_{i}$ with $\nu_{i}$. If the genus is equal to $1$, this
ensures that no higher powers of ${z\over \mu_{i}}$ and
${z\over \nu_{i}}$ are needed in the argument of the exponential.
However, there is a very simple
relation between $\mu_{i}$ and $\nu_{i}$. As remarked already in Sec. III,
the zeros of $F_{1}(z)$ are minus the zeros of $F_{2}(z)$ :
$\mu_{i}=-\nu_{i}$, $\forall i$. Hence
$$
F(z)={\widetilde \gamma} z^{2(n+1)}e^{(g_{1}+g_{2})(z)}
\prod_{i=1}^{\infty}\left(1-{z^{2}\over \mu_{i}^{2}}\right)
\; \; \; \; ,
\eqno (4.4a)
$$
where ${\widetilde \gamma}=\gamma_{1}\gamma_{2}$, and $\mu_{i}^{2}$
are the positive zeros of $F(z)$.

It turns out that the function
$(g_{1}+g_{2})$ in Eq. (4.4a) is actually a constant, so that we can write
$$
F(z)=F(-z)=
\gamma z^{2(n+1)}\prod_{i=1}^{\infty}\left(1-{z^{2}\over \mu_{i}^{2}}
\right) \; \; \; \; .
\eqno (4.4b)
$$
In fact, the following theorem holds (see Ref. 35, pp 250-251, and
in particular Ref. 36, pp 12-17).
\vskip 0.3cm
\noindent
{\bf Theorem 4.1} Let $f$ be an entire function. If $\forall \epsilon >0$
$\exists \; A_{\epsilon}$ such that :
$$
\log \max \Bigr \{1, \mid f(z) \mid \Bigr \} \leq A_{\epsilon}
{\mid z \mid}^{1+ \epsilon} \; \; \; \; ,
\eqno (4.5)
$$
then $f$ can be expressed in terms of its zeros as
$$
f(z)=e^{A+Bz}\prod_{i=1}^{\infty}\left(1-{z\over \widetilde \nu_{i}}\right)
e^{z\over \widetilde \nu_{i}} \; \; \; \; .
\eqno (4.6)
$$
If we now apply theorem 4.1 to the functions $F_{1}(z)z^{-(n+1)}$ and
$F_{2}(z)z^{-(n+1)}$ (see Eqs. (4.2,3)), we discover that the well-known
formula (see Ref. 32, relation 9.1.21, p 360) :
$$
J_{n}(z)={i^{-n}\over \pi}\int_{0}^{\pi}
e^{iz \cos \theta}\cos (n\theta) \; d \theta \; \; \; \; ,
\eqno (4.7)
$$
leads to the fulfillment of Eq. (4.5) for $F_{1}(z)z^{-(n+1)}$ and
$F_{2}(z)z^{-(n+1)}$. Thus these functions will satisfy Eq. (4.6) with
constants $A_{1}$ and $B_{1}$ for $F_{1}(z)z^{-(n+1)}$, and constants
$A_{2}$ and $B_{2}=-B_{1}$ for $F_{2}(z)z^{-(n+1)}$. The fact that
$B_{2}=-B_{1}$ is well-understood if we look again at Eqs. (4.2,3).
We are most indebted to Dr. R. Pinch for providing this argument.

The property that $F(z)$ admits the canonical-product expansion (4.4b) in
terms of eigenvalues permits us to evaluate $\zeta(0)$, where $\zeta(s)$
has been defined in Eq. (2.36), following a method described in Ref. 4.
The heat kernel $G(t)$, defined by
$$
G(t)\equiv \sum_{n,k}e^{-{\mid \lambda_{n,k} \mid}^{2}t}
\; \; \; \; ,
\eqno (4.8)
$$
has the familiar asymptotic expansion
$$
G(t)\sim \sum_{n=0}^{\infty}B_{n}t^{{n\over 2}-2} \; \; \; \; .
\eqno (4.9)
$$
valid as $t \rightarrow 0^{+}$, where in particular $B_{4}=\zeta(0)$.
Defining the generalized zeta-function
$$
\zeta(s,x^{2})\equiv\sum_{n,k}{\Bigr({\mid \lambda_{n,k} \mid}^{2}
+x^{2}\Bigr)}^{-s}
\; \; \; \; ,
\eqno (4.10)
$$
one then has
$$ \eqalignno{
\Gamma(3)\zeta(3,x^{2})&=\int_{0}^{\infty}t^{2}e^{-x^{2}t}G(t) \; dt \cr
&\sim \sum_{n=0}^{\infty}B_{n}\Gamma \left(1+{n\over 2}\right)x^{-2-n}
\; \; ,
& (4.11)\cr}
$$
where the asymptotic expansion holds as $x \rightarrow \infty$.
On the other hand, defining $m\equiv n+2$, one also has the identity
$$
\Gamma(3)\zeta(3,x^{2})=\sum_{m=2}^{\infty}N_{m}
{\left({1\over 2x}{d\over dx}\right)}^{3}
\log \Bigr[(ix)^{-2(m-1)}(J_{m-1}^{2}(ix)-J_{m}^{2}(ix))\Bigr] \; \; \; \; ,
\eqno (4.12)
$$
following from the canonical-product expansion (4.4b), where $N_{m}$ is the
degeneracy of the eigenvalues, given by $N_{m}=\Bigr(m^{2}-m\Bigr)$.
Since the functions $\Bigr(J_{m-1}^{2}(ix)-J_{m}^{2}(ix)\Bigr)$
admit a uniform Debye expansion at large
$x$,$^{32}$ one can obtain an asymptotic expansion of the right-hand side
of Eq. (4.12) valid as $x \rightarrow \infty$, which then yields an
alternative asymptotic expansion
$$
\Gamma(3)\zeta(3,x^{2})\sim \sum_{n=0}^{\infty}q_{n}x^{-2-n} \; \; \; \; ,
\eqno (4.13)
$$
valid as $x \rightarrow \infty$. By comparison of Eqs. (4.11) and (4.13) one
finds the coefficients $B_{n}$ and in particular $\zeta(0)=B_{4}=
{q_{4}\over 2}$.

The relevant uniform asymptotic expansions of $J_{n}(ix)$ and
$J_{n}'(ix)$, valid for $x \rightarrow \infty$ uniformly in the order $n$,
are given in Appendix A. The form of these expansions makes it necessary
to re-express the function $F(z)$ of Eq. (4.1) in terms of Bessel functions
and their derivatives of the same order. Using the identity
$$
J_{l}'(x)=J_{l-1}(x)-{l\over x}J_{l}(x) \; \; \; \; ,
\eqno (4.14)
$$
we find
$$ \eqalignno{
J_{m-1}^{2}(x)-J_{m}^{2}(x)&=\left(J_{m}'+{m\over x}J_{m}-J_{m}\right)
\left(J_{m}'+{m\over x}J_{m}+J_{m}\right) \cr
&={J_{m}'}^{2}+\left({m^{2}\over x^{2}}-1\right)J_{m}^{2}+
2{m\over x}J_{m}J_{m}' \; \; \; \; .
&(4.15)\cr}
$$
Thus, making the analytic continuation $x \rightarrow ix$,
defining $\alpha \equiv \sqrt{m^{2}+x^{2}}$ and using the notation of Eqs.
(A1,2), we obtain :
$$
J_{m-1}^{2}(ix)-J_{m}^{2}(ix)\sim {(ix)^{2(m-1)}\over 2\pi}
\alpha e^{2\alpha}e^{-2m \log(m+\alpha)}
\left[\Sigma_{1}^{2}+\Sigma_{2}^{2}+2{m\over \alpha}\Sigma_{1}\Sigma_{2}
\right] \; \; ,
\eqno (4.16)
$$
where the functions $\Sigma_{1}$ and $\Sigma_{2}$ have asymptotic series
as described in Appendix A :
$$
\Sigma_{1} \sim \sum_{k=0}^{\infty}{u_{k}({m\over \alpha})\over
m^{k}} \; \; \; \; ,
\eqno (4.17)
$$
$$
\Sigma_{2} \sim \sum_{k=0}^{\infty}{v_{k}({m\over \alpha})\over
m^{k}} \; \; \; \; ,
\eqno (4.18)
$$
and the functions $u_{k}$ and $v_{k}$ are polynomials given in Refs. 32, 37.
The asymptotic series on the right-hand sides of Eqs. (4.17,18) can be
re-expressed, defining
$$
t \equiv {m\over \alpha} \; \; \; \; ,
\eqno (4.19)
$$
as
$$
\sum_{k=0}^{\infty}{u_{k}({m\over \alpha})\over m^{k}} \sim
1+{a_{1}(t)\over \alpha}+{a_{2}(t)\over \alpha^{2}}+{a_{3}(t)\over
\alpha^{3}} + ... \; \; \; \; ,
\eqno (4.20)
$$
$$
\sum_{k=0}^{\infty}{v_{k}({m\over \alpha})\over m^{k}} \sim
1+{b_{1}(t)\over \alpha}+{b_{2}(t)\over \alpha^{2}}+
{b_{3}(t)\over \alpha^{3}}+... \; \; \; \; ,
\eqno (4.21)
$$
where
$$
a_{i}(t)={u_{i}(t)\over t^{i}}
\; \; \; \; , \; \; \; \;
b_{i}(t)={v_{i}(t)\over t^{i}}
\; \; \; \; , \; \; \; \; \forall i \geq 0 \; \; \; \; .
\eqno (4.22)
$$

Following Eqs. (4.1,12,16), we define
$$
{\widetilde \Sigma} \equiv \Sigma_{1}^{2}+\Sigma_{2}^{2}
+2t \Sigma_{1} \Sigma_{2} \; \; \; \; ,
\eqno (4.23)
$$
and study the asymptotic expansion of $\log({\widetilde \Sigma})$ in the
relation
$$
\log \left[(ix)^{-2(m-1)}\left(J_{m-1}^{2}-J_{m}^{2}\right)(ix)\right]
\sim -\log(2\pi)+ \log(\alpha) +2\alpha -2m \log(m+\alpha) +
\log ({\widetilde \Sigma}).
\eqno (4.24)
$$
{}From the relations (4.17)-(4.19) and (4.23), ${\widetilde \Sigma}$ has the
asymptotic expansion
$$
{\widetilde \Sigma} \sim c_{0}+{c_{1}\over \alpha}+{c_{2}\over \alpha^{2}}
+{c_{3}\over \alpha^{3}}+... \; \; \; \; ,
\eqno (4.25)
$$
where
$$
c_{0}=2(1+t) \; \; \; \; , \; \; \; \; \; \; \; \; c_{1}=2(1+t)(a_{1}+b_{1})
\; \; \; \; ,
\eqno (4.26,27)
$$
$$
c_{2}=a_{1}^{2}+b_{1}^{2}+2(1+t)(a_{2}+b_{2})+2ta_{1}b_{1} \; \; \; \; ,
\eqno (4.28)
$$
$$
c_{3}=2(1+t)(a_{3}+b_{3})+2(a_{1}a_{2}+b_{1}b_{2})+2t(a_{1}b_{2}+a_{2}b_{1})
\; \; \; \; .
\eqno (4.29)
$$
Higher-order terms have not been computed in Eq. (4.25) because they do not
affect the result for $\zeta(0)$, as we will show in detail in Sec. VII.
Defining
$$
\Sigma \equiv {{\widetilde \Sigma}\over c_{0}} \; \; \; \; ,
\eqno (4.30)
$$
and making the usual expansion
$$
\log(1+\omega)= \omega-{{\omega}^{2}\over 2}+{{\omega}^{3}\over 3}
-{{\omega}^{4}\over 4}
+{{\omega}^{5}\over 5}+... \; \; \; \; ,
\eqno (4.31)
$$
valid as $\omega \rightarrow 0$, we find
$$ \eqalignno{
\log({\widetilde \Sigma})&=\log(c_{0})+\log(\Sigma) \cr
& \sim \log(c_{0})+{A_{1}\over \alpha}+{A_{2}\over \alpha^{2}}+
{A_{3}\over \alpha^{3}}+... \; \; \; \; ,
& (4.32)\cr}
$$
where
$$
A_{1}=\left({c_{1}\over c_{0}}\right) \; \; \; \; , \; \; \; \; \; \; \; \;
A_{2}=\left({c_{2}\over c_{0}}\right)-
{{\left({c_{1}\over c_{0}}\right)}^{2}\over 2} \; \; \; \; ,
\eqno (4.33,34)
$$
$$
A_{3}=\left({c_{3}\over c_{0}}\right)-\left({c_{1}\over c_{0}}\right)
\left({c_{2}\over c_{0}}\right)+
{{\left({c_{1}\over c_{0}}\right)}^{3}\over 3} \; \; \; \; .
\eqno (4.35)
$$
Using Eqs. (2.13)-(2.23) of Ref. 37 and our Eqs. (4.22), (4.26)-(4.29),
(4.33)-(4.35), we find after a lengthy calculation :
$$
A_{1}=\sum_{r=0}^{2}k_{1r}t^{r} \; \; \; \; , \; \; \; \; \; \; \; \;
A_{2}=\sum_{r=0}^{4}k_{2r}t^{r}
\; \; \; \; , \; \; \; \; \; \; \; \;
A_{3}=\sum_{r=0}^{6}k_{3r}t^{r} \; \; \; \; ,
\eqno (4.36)
$$
where
$$
k_{10}=-{1\over 4} \; \; \; \; , \; \; \; \; k_{11}=0 \; \; \; \; , \; \;
\; \; k_{12}={1\over 12} \; \; \; \; ,
\eqno (4.37)
$$
$$
k_{20}=0 \; \; \; \; , \; \; \; \; k_{21}=-{1\over 8} \; \; \; \; , \; \;
\; \; k_{22}=k_{23}={1\over 8} \; \; \; \; , \; \; \; \;
k_{24}=-{1\over 8} \; \; \; \; ,
\eqno (4.38)
$$
$$
k_{30}={5\over 192} \; \; \; \; , \; \; \; \;
k_{31}=-{1\over 8} \; \; \; \; , \; \; \; \;
k_{32}={9\over 320} \; \; \; \; , \; \; \; \;
k_{33}={1\over 2} \; \; \; \; ,
\eqno (4.39a)
$$
$$
k_{34}=-{23\over 64} \; \; \; \; , \; \; \; \;
k_{35}=-{3\over 8} \; \; \; \; , \; \; \; \;
k_{36}={179\over 576} \; \; \; \; .
\eqno (4.39b)
$$
Note that, interestingly, all factors of $(1+t)$ in the denominators of
$A_{1},A_{2}$ and $A_{3}$ in Eqs. (4.33)-(4.35) have cancelled against
factors in the numerators.

The result of these calculations, following from Eqs. (4.24) and
(4.32)-(4.39), may be summarized in the expression
$$
\log \left[(ix)^{-2(m-1)}\left(J_{m-1}^{2}-J_{m}^{2}\right)(ix)\right]
\sim \sum_{i=1}^{5}S_{i}(m,\alpha(x))+ \; higher-order \; terms,
\eqno (4.40)
$$
where
$$
S_{1}\equiv -\log(\pi)+2\alpha \; \; \; \; ,
\eqno (4.41)
$$
$$
S_{2}\equiv -(2m-1)\log(m+\alpha) \; \; \; \; ,
\eqno (4.42)
$$
$$
S_{3}\equiv \sum_{r=0}^{2}k_{1r}m^{r}\alpha^{-r-1} \; \; \; \; ,
\eqno (4.43)
$$
$$
S_{4}\equiv \sum_{r=0}^{4}k_{2r}m^{r}\alpha^{-r-2} \; \; \; \; ,
\eqno (4.44)
$$
$$
S_{5}\equiv \sum_{r=0}^{6}k_{3r}m^{r}\alpha^{-r-3} \; \; \; \; .
\eqno (4.45)
$$
The asymptotic series (4.40) will be sufficient for computing $\zeta(0)$.
The infinite sum over $m$ in the expression (4.12) can be evaluated
following Eq. (4.40) with the help of the formulae derived using contour
integration : $^{4}$
$$
\sum_{m=0}^{\infty}m^{2k}\alpha^{-2k-l}=
{\Gamma \Bigr(k+{1\over 2}\Bigr)\Gamma \Bigr({l\over 2}-{1\over 2}\Bigr)
\over
2\Gamma \Bigr(k+{l\over 2}\Bigr)}x^{1-l} \; \; , \; \;
k=1,2,3,... \; \; \; \; ,
\eqno (4.46)
$$
$$
\sum_{m=0}^{\infty}m \alpha^{-1-l} \sim {x^{1-l}\over \sqrt{\pi}}
\sum_{r=0}^{\infty}{2^{r}\over r!}{\widetilde B}_{r}x^{-r}
{\Gamma \Bigr({r\over 2}+{1\over 2}\Bigr)
\Gamma \Bigr({l\over 2}-{1\over 2}+{r\over 2}\Bigr) \over
2\Gamma \Bigr({1\over 2}+{l\over 2}\Bigr)}
\cos \Bigr({r \pi \over 2}\Bigr) \; \; \; \; .
\eqno (4.47)
$$
Here $l$ is a real number obeying $l>1$ and ${\widetilde B}_{0}=1,
{\widetilde B}_{2}={1\over 6}, {\widetilde B}_{4}=-{1\over 30}$ etc. are
Bernoulli numbers. Thus, defining the functions $W_{\infty}$ and
$W_{\infty}^{1},...,W_{\infty}^{5}$ of $x$ by
$$
W_{\infty}\equiv \sum_{m=0}^{\infty}\left(m^{2}-m\right)
{\left({1\over 2x}{d\over dx}\right)}^{3}\left[\sum_{i=1}^{5}
S_{i}(m,\alpha(x))\right]=\sum_{i=1}^{5}W_{\infty}^{i} \; \; ,
\eqno (4.48)
$$
we find following Eq. (4.12) and the discussion of Sec. VII that
$$
\Gamma(3)\zeta(3,x^{2})
\sim W_{\infty}+\sum_{n=5}^{\infty}{\hat q}_{n}x^{-2-n} \; \; \; \; .
\eqno (4.49)
$$
Since, in Eq. (4.11), $\zeta(0)=B_{4}$ is found from the coefficient of
$x^{-6}$, we see that $\zeta(0)$ can be computed from the asymptotic
series for $W_{\infty}$ as $x \rightarrow \infty$.
\vskip 0.3cm
\centerline {\bf V. CONTRIBUTION OF $W_{\infty}^{1}$
AND $W_{\infty}^{2}$}
\vskip 0.3cm
The term $W_{\infty}^{1}$ does not contribute to $\zeta(0)$. In fact, using
Eqs. (4.41) and (4.46)-(4.48) we find
$$ \eqalignno{
W_{\infty}^{1}&={3\over 4}\sum_{m=0}^{\infty}\left(m^{2}-m\right)\alpha^{-5}
\cr
&\sim {x^{-2}\over 4}
-{3\over 4}{x^{-3}\over \Gamma({1\over 2})}
\sum_{r=0}^{\infty}{2^{r}\over r!}{\widetilde B}_{r}x^{-r}
{\Gamma\left({r\over 2}+{1\over 2}\right)
\Gamma\left({r\over 2}+{3\over 2}\right)\over
2\Gamma \left({5\over 2}\right)}
\cos \left({r\pi \over 2}\right) \; \; \; \; ,
&(5.1)\cr}
$$
which implies that $x^{-6}$ (whose coefficient contributes to $\zeta(0)$)
does not appear.

The contribution $\zeta^{2}(0)$ of $W_{\infty}^{2}$ to $\zeta(0)$ is
obtained using the identities
$$
{\left({1\over 2x}{d\over dx}\right)}^{3}\log
\left({1\over {m+\alpha}}\right)=(m+\alpha)^{-3}
\left[-\alpha^{-3}-{9\over 8}m\alpha^{-4}-{3\over 8}m^{2}\alpha^{-5}\right]
\; \; ,
\eqno (5.2)
$$
$$
(m+\alpha)^{-3}={(\alpha -m)^{3}\over x^{6}} \; \; \; \; ,
\eqno (5.3)
$$
so that Eq. (5.2) becomes
$$
{\left({1\over 2x}{d\over dx}\right)}^{3}\log
\left({1\over {m+\alpha}}\right)=
-x^{-6}+mx^{-6}\alpha^{-1}+{m\over 2}x^{-4}\alpha^{-3}
+{3\over 8}mx^{-2}\alpha^{-5} \; \; \; \; .
\eqno (5.4)
$$
Now the series for $W_{\infty}^{2}$ is convergent, as may be checked using
Eqs. (4.42,48) and the form (5.2). However, when the sum over $m$ is
rewritten using the splitting (5.4), the individual pieces become
divergent. These `fictitious' divergences may be regularized using the
device of Ref. 4 : dividing by $\alpha^{2s}$, summing using Eqs. (4.46,47)
and then taking the limit $s\rightarrow 0$.
With this understanding, and using the sums $\rho_{i}$
defined in Eqs. (B1)-(B12), we find :
$$ \eqalignno{
W_{\infty}^{2}&=-2x^{-6}\rho_{1}+2x^{-6}\rho_{2}+x^{-4}\rho_{3}
+{3\over 4}x^{-2}\rho_{4}+3x^{-6}\rho_{5}-3x^{-6}\rho_{6}\cr
&-{3\over 2}x^{-4}\rho_{7}-{9\over 8}x^{-2}\rho_{8}-x^{-6}\rho_{9}
+x^{-6}\rho_{10}+{x^{-4}\over 2}\rho_{11}+{3\over 8}x^{-2}\rho_{12}
\; \; .
&(5.5)\cr}
$$
When odd powers of $m$ greater than $1$ occur, we can still apply Eq. (4.47)
after re-expressing $m^{2}$ as $\alpha^{2}-x^{2}$. Thus,
applying again the contour formulae (4.46,47), only $\rho_{1}$
and $\rho_{9}$ are found to contribute to $\zeta^{2}(0)$, leading to
$$
\zeta^{2}(0)=-{1\over 120}+{1\over 24}={1\over 30} \; \; \; \; .
\eqno (5.6)
$$
\vskip 0.3cm
\centerline {\bf VI. EFFECT OF $W_{\infty}^{3}, W_{\infty}^{4}$ AND
$W_{\infty}^{5}$}
\vskip 0.3cm
The term $W_{\infty}^{3}$ does not contribute to $\zeta(0)$. In fact,
using the relations (4.43) and (4.48) we find
$$
W_{\infty}^{3}=-{1\over 8}\sum_{r=0}^{2}k_{1r}(r+1)(r+3)(r+5)
\left[\sum_{m=0}^{\infty}\left(m^{2}-m\right)m^{r}\alpha^{-r-7}\right]
\; \; .
\eqno (6.1)
$$
In light of Eqs. (4.46,47), $x^{-6}$ does not appear in the asymptotic
expansion of Eq. (6.1) at large $x$.

For the term $W_{\infty}^{4}$ a remarkable cancellation occurs. In fact,
using Eqs. (4.44,48) we find
$$
W_{\infty}^{4}=-{1\over 8}\sum_{r=0}^{4}k_{2r}(r+2)(r+4)(r+6)
\left[\sum_{m=0}^{\infty}\left(m^{2}-m\right)m^{r}\alpha^{-r-8}\right]
\; \; .
\eqno (6.2)
$$
The application of Eqs. (4.38,46,47) leads to
$$
\zeta^{4}(0)={1\over 2}\sum_{r=0}^{4}k_{2r}=0 \; \; \; \; .
\eqno (6.3)
$$
Finally, using Eqs. (4.45,48) we find
$$
W_{\infty}^{5}=-{1\over 8}\sum_{r=0}^{6}k_{3r}(r+3)(r+5)(r+7)
\left[\sum_{m=0}^{\infty}\left(m^{2}-m\right)m^{r}\alpha^{-r-9}\right]
\; \; .
\eqno (6.4)
$$
Again, the contour formulae (4.46,47) lead to
$$
\zeta^{5}(0)=-{1\over 2}\sum_{r=0}^{6}k_{3r}=-{1\over 360} \; \; \; \; ,
\eqno (6.5)
$$
in light of Eq. (4.39).
\vskip 10cm
\centerline {\bf VII. VANISHING EFFECT OF HIGHER-ORDER TERMS}
\vskip 1cm
We now prove the statement made after Eq. (4.29), namely
that there is no need to compute the explicit form of $c_{k}$ in Eq. (4.25),
$\forall k>3$. In fact, the formulae (4.33)-(4.35) can be completed by
$$
A_{4}=\left({c_{4}\over c_{0}}\right)
-{{\left({c_{2}\over c_{0}}\right)}^{2}\over 2}
-\left({c_{1}\over c_{0}}\right)\left({c_{3}\over c_{0}}\right)
+{\left({c_{1}\over c_{0}}\right)}^{2}
\left({c_{2}\over c_{0}}\right)
-{{\left({c_{1}\over c_{0}}\right)}^{4}\over 4} \; \; ,
\eqno (7.1)
$$
plus infinitely many others, and the general term has the structure
$$
A_{n}=\sum_{p=1}^{l}h_{np}(1+t)^{-p}+\sum_{r=0}^{2n}k_{nr}t^{r} \; \; \; \; ,
\; \; \; \; \forall n \geq 1 \; \; \; \; ,
\eqno (7.2)
$$
where $l<n$, the $h_{np}$ are constants, and
$r$ assumes both odd and even values.
We have indeed proved that $h_{11}=h_{21}=h_{31}=0$, but the calculation
of $h_{np}$ for all values of $n$ is not obviously feasible.
However, we will show that the exact
value of $h_{np}$ does not affect the $\zeta(0)$ value.
Thus, $\forall n>3$, we must study
$$
H_{\infty}^{n}\equiv \sum_{m=0}^{\infty}\left(m^{2}-m\right)
{\left({1\over 2x}{d\over dx}\right)}^{3}\left[{A_{n}\over \alpha^{n}}\right]
=H_{\infty}^{n,A}+H_{\infty}^{n,B} \; \; \; \; ,
\eqno (7.3)
$$
where, defining
$$
a_{np}\equiv (p-n)(p-n-2)(p-n-4) \; , \;
b_{np}\equiv 3\Bigr(-p^{3}+(3+2n)p^{2}-(n^{2}+3n+1)p \Bigr) \; ,
\eqno (7.4a)
$$
$$
c_{np}\equiv 3\Bigr(p^{3}-\left(p^{2}+p\right)n-p \Bigr)
\; \; \; \; , \; \; \; \;
d_{np}\equiv -p(p+1)(p+2) \; \; \; \; ,
\eqno (7.4b)
$$
one has
$$ \eqalignno{
H_{\infty}^{n,A}&\equiv \sum_{p=1}^{l}h_{np}
\sum_{m=0}^{\infty}\left(m^{2}-m\right)
{\left({1\over 2x}{d\over dx}\right)}^{3}\Bigr[\alpha^{p-n}
(m+\alpha)^{-p}\Bigr]\cr
&=\sum_{p=1}^{l}{h_{np}\over 8}\sum_{m=0}^{\infty}\left(m^{2}-m\right)
\Bigr[a_{np}\alpha^{p-n-6}(m+\alpha)^{-p}+b_{np}\alpha^{p-n-5}
(m+\alpha)^{-p-1}  \cr
& +c_{np}\alpha^{p-n-4}(m+\alpha)^{-p-2}
+d_{np}\alpha^{p-n-3}(m+\alpha)^{-p-3}\Bigr] \; \; \; \; ,
&(7.5)\cr}
$$
$$
H_{\infty}^{n,B}\equiv -{1\over 8}\sum_{r=0}^{2n}k_{nr}(r+n)(r+n+2)
(r+n+4)\left[\sum_{m=0}^{\infty}\left(m^{2}-m\right)m^{r}
\alpha^{-r-n-6}\right].
\eqno (7.6)
$$
Because we are only interested in understanding the behaviour of Eq. (7.5)
as a function of $x$, the application of the Euler-Maclaurin formula$^{38}$
described in Appendix C
is more useful than the splitting (5.3). In so doing, we find that
the part of the Euler-Maclaurin formula involving the integral on the
left-hand side of Eq. (C2), when $n=\infty$, contains the least negative power
of $x$. Thus, if we prove that the conversion of Eq. (7.5) into an integral
only contains $x^{-l}$ with $l>6, \; \; \forall n>3$, we have proved that
$H_{\infty}^{n,A}$ does not contribute to $\zeta(0)$, $\forall n>3$. This is
indeed the case, because in so doing we deal with the integrals defined
in Eqs. (C4)-(C11),
where $I_{1}^{np}, I_{3}^{np}, I_{5}^{np}$ and $I_{7}^{np}$
are proportional to $x^{-3-n}$, and $I_{2}^{np}, I_{4}^{np}, I_{6}^{np}$
and $I_{8}^{np}$ are proportional to $x^{-4-n}$, where $n>3$.

Finally, in Eq. (7.6) we must study the case when $r$ is even and the case
when $r$ is odd. In so doing, defining
$$
\Sigma_{(I)}\equiv \sum_{m=0}^{\infty}m^{2+r}\alpha^{-r-n-6}
\; \; \; \; , \; \; \; \; \; \; \; \;
\Sigma_{(II)}\equiv \sum_{m=0}^{\infty}m^{1+r}\alpha^{-r-n-6}
\; \; \; \; ,
\eqno (7.7)
$$
we find for $r=2k>0$ ($k$ integer)
$$
\Sigma_{(I)}= {x^{-3-n}\over 2}
{\Gamma \left(k+{3\over 2}\right) \Gamma \left({n\over 2}+{3\over 2}\right)
\over
\Gamma \left(3+k+{n\over 2}\right)} \; \; \; \; ,
\eqno (7.8)
$$
and for $r=2k+1 \; \;$ ($k$ integer $\geq 0$)
$$ \eqalignno{
\Sigma_{(I)}&\sim
{x^{-3-n}\over \Gamma \left({1\over 2}\right)}
\sum_{l=0}^{\infty} \left \{{2^{l}\over l!}{{\widetilde B}_{l}\over 2}x^{-l}
\Gamma \left({l\over 2}+{1\over 2}\right)\cos \left({l \pi \over 2}\right)
\right. \cr
&\left. \left[ {\Gamma \left({n\over 2}+{3\over 2}+{l\over 2}\right)
\over \Gamma \left({n\over 2}+{5\over 2}\right)}+...+(-1)^{(1+k)}
{\Gamma \left({n\over 2}+{5\over 2}+k +{l\over 2}\right) \over
\Gamma \left({n\over 2}+{7\over 2}+k \right)}\right]\right \} \; \; .
&(7.9)\cr}
$$
Moreover, we find for $r=2k>0$
$$ \eqalignno{
\Sigma_{(II)}&\sim
{x^{-4-n}\over \Gamma \left({1\over 2}\right)}\sum_{l=0}^{\infty}
\left \{{2^{l}\over l!}{{\widetilde B}_{l}\over 2}x^{-l}
\Gamma \left({l\over 2}+{1\over 2}\right)\cos \left({l\pi \over 2}\right)
\right. \cr
&\left. \left[ {\Gamma \left({n\over 2}+2+{l\over 2}\right) \over
\Gamma \left({n\over 2}+3 \right)}+... +(-1)^{k}
{\Gamma \left({n\over 2}+2+k+{l\over 2}\right)\over
\Gamma \left({n\over 2}+3+k \right)}\right]\right \} \; \; ,
&(7.10)\cr}
$$
and for $r=2k+1 \; \; (k \geq 0)$
$$
\Sigma_{(II)}= {x^{-4-n}\over 2}
{\Gamma \left(k+{3\over 2}\right)\Gamma \left({n\over 2}+2\right) \over
\Gamma \left({7\over 2}+k+{n\over 2}\right)} \; \; \; \; .
\eqno (7.11)
$$
Once more, in deriving Eqs. (7.8,11) we used Eq. (4.46), and in deriving
Eqs. (7.9,10) we used Eq. (4.47). Thus also $H_{\infty}^{n,B}$ does not
contribute to $\zeta(0), \; \; \forall n>3$, and our proof is completed.
\vskip 10cm
\centerline {\bf VIII. $\zeta(0)$ VALUE}
\vskip 1cm
In light of Eqs. (5.6, 6.3, 6.5), and using the result proved in Sec. VII,
we conclude that for the complete Majorana field $\Bigr(\psi^{A}, \;
{\widetilde \psi}^{A'}\Bigr)$ :
$$
\zeta(0)={11\over 360} \; \; \; \; .
\eqno (8.1)
$$
This value, found by a {\it direct} calculation, disagrees with the result
$\zeta(0)={17\over 180}$ in Ref. 11, which was computed using a
different, {\it indirect}, approach. If there are $N$ massless Majorana
fields, the full $\zeta(0)$ in Eq. (8.1) should be multiplied by $N$.
Following the remarks in the Introduction, this $\zeta(0)$ value can now be
combined with the $\zeta(0)$ values for other spins, in order to check
whether the one-loop divergences in quantum cosmology cancel in
higher-$N$ supergravity theories (in the case of our simple background
geometry). At present it can only be said that existing results are
inconclusive on this question (see Sec. IX).
\vskip 1cm
\centerline {\bf IX. CONCLUDING REMARKS}
\vskip 1cm
Note, following the comments at the end of Sec. II, that the zeta-function
studied in Ref. 11, formed from the eigenvalues of the squared-up
Dirac operator, is the same as the zeta-function studied here, formed from
squared eigenvalues, since the boundary conditions agree. Unfortunately it
is hard to understand the discrepancy between the spin-${1\over 2}$ result
of Ref. 11 and the present result, since the approach of Refs. 9, 11
involves finding general expressions for $\zeta(0)$ for
various spins and types of boundary condition, on a general Riemannian
manifold with boundary. Only at the end of the calculation in Ref. 11
does one restrict attention to a flat background with spherical boundary;
degeneracies and a specific eigenvalue equation play no role.

Turning to the question of $\zeta(0)$ in extended supergravity models with
boundaries present, recall that two sets of possible `supersymmetric'
boundary conditions were suggested in the Introduction. The first set,
Breitenlohner-Freedman-Hawking boundary conditions,$^{20,21}$ involve field
strengths at the boundary for gauge fields (spins $s=1,{3\over 2},2$), being
either of magnetic or electric type, depending on the sign of the quantity
$\epsilon$ introduced in Sec. I. For bosonic fields, the magnetic conditions
involve fixing the magnetic field $B_{i}$ on the surface ($s=1$) or fixing
the magnetic part $B_{ij}$ of the Weyl tensor ($s=2$), which involves the
normal derivative of the metric. Similarly the electric conditions fix the
electric field $E_{i}$ ($s=1$) or the electric part $E_{ij}$ of the Weyl
tensor ($s=2$). This points to a difficulty with such boundary conditions,
since $E_{ij}$ involves second derivatives of the metric in the normal
direction, so that electric boundary conditions cannot be formulated in
terms of canonical variables (using Hamiltonian language) for $s=2$. This
in turn affects the quantization of such a system, since in the case when
$E_{ij}$ is fixed one cannot find a Euclidean action $I$ for linearized
gravity with these boundary conditions, such that $\delta I=0$ gives the
linearized $s=2$ field equations subject to specified $E_{ij}$ at the
boundary. As a result, one cannot formulate a Hartle-Hawking path integral
in the usual way. A similar problem arises for $s={3\over 2}$, in both the
magnetic and electric cases : the boundary conditions again cannot be
written in terms of canonical variables, so that again one cannot write
down the quantum amplitude as a path integral. Thus the
Breitenlohner-Freedman-Hawking boundary conditions are problematic for
spins $s={3\over 2}$ and $2$ in our example.

One is therefore led to study the alternative {\it locally supersymmetric}
boundary conditions suggested in the Introduction, following
from Ref. 24, and used in Ref. 11. In the case of a
spin-${1\over 2}$ field, these again take the form (1.1), and for a
complex scalar field they again involve Dirichlet and Neumann conditions
on the real and imaginary parts (or viceversa). For spin $1$, they can be
regarded as magnetic, and for spins ${3\over 2}$ and $2$ they involve
fixing the projection $\Bigr(\epsilon \; {\widetilde \psi}_{i}^{A'}
-\sqrt{2} \; {_{e}n_{A}^{\; \; A'}}\psi_{i}^{A}\Bigr)$ of the
spin-${3\over 2}$ potential and the spatial tetrad $e_{\; \; \; \; \;i}^{AA'}$
(and hence intrinsic metric $h_{ij}$) up to gauge. Refs. 9, 11 give
detailed calculations of $\zeta(0)$ in a general background geometry for
these boundary conditions, including the contribution of gauge-fixing and
ghost terms. They find no cancellations when summing over spins for any
higher-$N$ supergravity theory, allowing also for the topological
contribution of anti-symmetric tensor fields where appropriate.$^{9,11}$
In the case of flat Euclidean four-space bounded by a three-sphere, the
resulting $\zeta(0)$ value for $s={1\over 2}$ has already been quoted. For
$s=0$ the result $\zeta(0)={7\over 45}$ can be confirmed by a {\it direct}
calculation based on finding the eigenvalue conditions and using the
methods of Ref. 4. The values found in Ref. 11 for $s=1,{3\over 2}$ and $2$
in this case are $\zeta(0)=-{38\over 45},{197\over 180}$ and
$-{803\over 45}$. However, the discrepancy found in this paper for
$s={1\over 2}$ makes it particularly necessary to check also the result for
$s={3\over 2}$; both fermionic cases $s={1\over 2}$ and ${3\over 2}$
involve mixed boundary conditions in a crucial way, and are somewhat similar,
although the extra technical difficulties of spin ${3\over 2}$ may render a
{\it direct} calculation of $\zeta(0)$ along the lines of this paper
impossible. Given these uncertainties, one cannot say that the question
of the one-loop finiteness of extended supergravity, in the presence of
boundaries, has been settled.

A further interesting question, related to recent considerations in the
literature,$^{39,40}$ is raised by this work in the case of gauge
fields. The
value of $\zeta(0)$ depends on whether one first restricts the classical
theory to a set of physical degrees of freedom, by choice of gauge, and then
quantizes, or whether one quantizes the full theory in BRST-invariant
fashion with gauge-averaging and ghost terms included. For example, the
original quantum gravity ($s=2$) calculation of Schleich$^{1}$ worked with
physical degrees of freedom, using a transverse-traceless gauge and
Dirichlet boundary conditions on the three-sphere to obtain
$\zeta(0)=-{278\over 45}$, which differs from the result
$\zeta(0)=-{803\over 45}$ of Ref. 11. Similarly the $s=1$ calculation of
Louko,$^{2}$ working in a transverse gauge with Dirichlet (magnetic) boundary
conditions on the three-sphere, gave $\zeta(0)=-{77\over 180}$, as compared
to the result $\zeta(0)=-{38\over 45}$ of Ref. 11. [Electric boundary
conditions give $\zeta(0)={13\over 180}$ with physical degrees of
freedom,$^{8}$ while again $\zeta(0)=-{38\over 45}$ in Ref. 11]. The present
authors have also done a calculation for spin ${3\over 2}$, working with
physical degrees of freedom, in the gauge $e_{AA'}^{\; \; \; \; \; j}
\psi_{j}^{A}=0, \; e_{AA'}^{\; \; \; \; \; j}{\widetilde \psi}_{j}^{A'}=0$,
subject to the boundary conditions of the previous
paragraph on the three-sphere, giving $\zeta(0)=-{289\over 360}$, as
compared to the result $\zeta(0)={197\over 180}$ of Ref. 11. In the BRST
method, the local boundary conditions are chosen such that
transition amplitudes are invariant under BRST transformations, and hence,
by the Fradkin-Vilkovisky theorem,$^{41}$ the amplitudes do not depend on
the gauge-fixing term. In a related example,$^{42}$ the method involving
physical degrees of freedom has been shown to be formally equivalent to a
method involving Faddeev-Popov ghosts (and hence to a BRST method). But
apparently in the present example of quantum fields inside a three-sphere,
the relation between the two approaches is only formal. One expects,
for example, that the $\zeta(0)$ value for a
physical-degrees-of-freedom calculation might depend on the choice of gauge
conditions. In this situation, since gauge invariance is the underlying
physical principle, one is forced to choose the gauge-invariant
BRST method.
\vskip 1cm
\centerline {\bf ACKNOWLEDGMENTS}
\vskip 1cm
We are very grateful to Gary Gibbons, Stephen Hawking, Jorma Louko, Richard
Pinch and Claudia Yastremiz for conversations, and to Hugh Luckock,
Ian Moss, Stephen Poletti and Kristin Schleich for correspondence and
conversations on our work.
\vskip 30cm
\centerline {\bf APPENDIX A}
\vskip 1cm
In Sec. IV, we use uniform asymptotic expansions of regular Bessel
functions $J_{n}$ and their first derivatives $J_{n}'$, relying
on the relations 9.3.7 and 9.3.11
on p 366 of Ref. 32. In those formulae,
the argument of $J_{n}$ and $J_{n}'$ is ${n\over \cosh \gamma}$, where
$\gamma$ is fixed and positive and $n$ is large and positive. If we put
${n\over \cosh \gamma}=x$, we find that
$$
e^{\gamma}={n\over x} \pm \sqrt{{n^{2}\over x^{2}}-1} \; \; \; \; ,
\; \; \; \;
\tanh \gamma=\pm {1\over n}\sqrt{n^{2}-x^{2}} \; \; \; \; .
$$
Thus, making the analytic continuation $x\rightarrow ix$ and then
defining $\alpha=\sqrt{n^{2}+x^{2}}$, we may write
$$
J_{n}(ix)\sim {(ix)^{n}\over \sqrt{2\pi}}{\alpha}^{-{1\over 2}}e^{\alpha}
e^{-n \log(n+\alpha)}\Sigma_{1} \; \; \; \; ,
\eqno (A1)
$$
$$
J_{n}'(ix)\sim {(ix)^{n-1}\over \sqrt{2\pi}}{\alpha}^{1\over 2}e^{\alpha}
e^{-n \log(n+\alpha)}\Sigma_{2} \; \; \; \; ,
\eqno (A2)
$$
where the functions $\Sigma_{1}$ and $\Sigma_{2}$ admit the asymptotic
expansions :
$\Sigma_{1} \sim \sum_{k=0}^{\infty}{u_{k}
({n\over \alpha}) \over n^{k}}$ ,
$\Sigma_{2} \sim \sum_{k=0}^{\infty}{v_{k}({n\over \alpha}) \over n^{k}}$,
valid uniformly in the order $n$ as $\mid x \mid \rightarrow \infty$.
The functions $u_{k}$ and $v_{k}$ are polynomials, given by the
relations (9.3.9) and (9.3.13) on p 366 of Ref. 32.
\vskip 10cm
\centerline {\bf APPENDIX B}
\vskip 1cm
We now write the infinite sums used for the $\zeta(0)$ calculation of
Secs. IV-VIII. In Eq. (5.5) we have
$$
\rho_{1}\equiv \sum_{m=0}^{\infty}m^{3} \; \; \; \; , \; \; \; \;
\rho_{2}\equiv \sum_{m=0}^{\infty}m^{4}\alpha^{-1}
\; \; \; \; , \; \; \; \;
\rho_{3}\equiv \sum_{m=0}^{\infty}m^{4}\alpha^{-3} \; \; \; \; ,
\eqno (B1,2,3)
$$
$$
\rho_{4}\equiv \sum_{m=0}^{\infty}m^{4}\alpha^{-5}
\; \; \; \; , \; \; \; \;
\rho_{5}\equiv \sum_{m=0}^{\infty}m^{2}
\; \; \; \; , \; \; \; \;
\rho_{6}\equiv \sum_{m=0}^{\infty}m^{3}\alpha^{-1} \; \; \; \; ,
\eqno (B4,5,6)
$$
$$
\rho_{7}\equiv \sum_{m=0}^{\infty}m^{3}\alpha^{-3}
\; \; \; \; , \; \; \; \;
\rho_{8}\equiv \sum_{m=0}^{\infty}m^{3}\alpha^{-5}
\; \; \; \; , \; \; \; \;
\rho_{9}\equiv \sum_{m=0}^{\infty}m \; \; \; \; ,
\eqno (B7,8,9)
$$
$$
\rho_{10}\equiv \sum_{m=0}^{\infty}m^{2}\alpha^{-1}
\; \; \; \; , \; \; \; \;
\rho_{11}\equiv \sum_{m=0}^{\infty}m^{2}\alpha^{-3}
\; \; \; \; , \; \; \; \;
\rho_{12}\equiv \sum_{m=0}^{\infty}m^{2}\alpha^{-5} \; \; \; \; .
\eqno (B10,11,12)
$$
Of course, the sums (B1)-(B12) do not by themselves make sense as they are
written. But, as discussed in Sec. V, the sums appear in convergent linear
combinations and their values are regularized as explained following
Eq. (5.4).
\vskip 10cm
\centerline {\bf APPENDIX C}
\vskip 1cm
If $f$ is a function which obeys suitable differentiability and
integrability conditions, the Euler-Maclaurin formula is a very useful tool
which can be used to estimate the sum $\sum_{i=0}^{n}f(i)$. Denoting by
${\widetilde B}_{s}$ the Bernoulli numbers, defined by the expansion
$$
{t\over e^{t}-1}=\sum_{s=0}^{\infty}{\widetilde B}_{s}{t^{s}\over s!}
\; \; \; \; , \; \; \; \; \mid t \mid < 2\pi \; \; \; \; ,
\eqno (C1)
$$
the following theorem holds.$^{38}$

{\bf Theorem C.1} Let $f$ be a real- or complex-valued function defined
on $0\leq t < \infty$. If $f^{(2m)}(t)$ is absolutely integrable on
$(0,\infty)$ then, for $n=1,2,...,$
$$ \eqalignno{
\sum_{i=0}^{n}f(i)-\int_{0}^{n}f(x) \; dx &=
{1\over 2}\Bigr[f(0)+f(n)\Bigr]\cr
&+ \sum_{s=1}^{m-1}{{\widetilde B}_{2s}\over (2s)!}\Bigr[f^{(2s-1)}(n)
-f^{(2s-1)}(0)\Bigr]+R_{m}(n) \; \; ,
&(C2)\cr}
$$
where the remainder $R_{m}(n)$ satisfies
$$
\mid R_{m}(n) \mid \; \leq \; \Bigr(2-2^{1-m}\Bigr)
{{\mid {\widetilde B}_{2m} \mid} \over (2m)!}
\int_{0}^{n} \mid {f^{(2m)}(x)} \mid \; dx \; \; \; \; .
\eqno (C3)
$$
Eq. (C2) can be used to evaluate infinite sums, setting $n=\infty$, if the
corresponding derivatives, and the integrals in Eqs. (C2,3) are well-defined.
Typically one considers ${\widetilde n}$ terms in the sum involving
Bernoulli numbers in Eq. (C2), and neglects terms starting with some value
$s={\widetilde n}+1 \; \leq \; (m-1)$. Eq. (C3) can
be used to show that the remainder $R_{m}(n)$ is bounded in absolute
value by a constant times the first neglected term in the sum
in Eq. (C2), provided
$f^{(2m)}$ obeys suitable conditions specified on p 38 of Ref. 38.

The integrals which arise in Sec. VII in taking $n=\infty$ on the left-hand
side of Eq. (C2) are
$$
I_{1}^{np}\equiv \int_{0}^{\infty}y^{2}
{\left(y+\sqrt{x^{2}+y^{2}}\right)}^{-p}
{\Bigr(x^{2}+y^{2}\Bigr)}^{{p\over 2}-{n\over 2}-3} \; dy \; \; \; \; ,
\eqno (C4)
$$
$$
I_{2}^{np}\equiv \int_{0}^{\infty}y
{\left(y+\sqrt{x^{2}+y^{2}}\right)}^{-p}
{\Bigr(x^{2}+y^{2}\Bigr)}^{{p\over 2}-{n\over 2}-3} \; dy  \; \; \; \; ,
\eqno (C5)
$$
$$
I_{3}^{np}\equiv \int_{0}^{\infty}y^{2}
{\left(y+\sqrt{x^{2}+y^{2}}\right)}^{-p-1}
{\Bigr(x^{2}+y^{2}\Bigr)}^{{p\over 2}-{n\over 2}-{5\over 2}} \; dy
\; \; \; \; ,
\eqno (C6)
$$
$$
I_{4}^{np}\equiv \int_{0}^{\infty}y
{\left(y+\sqrt{x^{2}+y^{2}}\right)}^{-p-1}
{\Bigr(x^{2}+y^{2}\Bigr)}^{{p\over 2}-{n\over 2}-{5\over 2}} \; dy
\; \; \; \; ,
\eqno (C7)
$$
$$
I_{5}^{np}\equiv \int_{0}^{\infty}y^{2}
{\left(y+\sqrt{x^{2}+y^{2}}\right)}^{-p-2}
{\Bigr(x^{2}+y^{2}\Bigr)}^{{p\over 2}-{n\over 2}-2} \; dy \; \; \; \; ,
\eqno (C8)
$$
$$
I_{6}^{np}\equiv \int_{0}^{\infty}y
{\left(y+\sqrt{x^{2}+y^{2}}\right)}^{-p-2}
{\Bigr(x^{2}+y^{2}\Bigr)}^{{p\over 2}-{n\over 2}-2} \; dy \; \; \; \; ,
\eqno (C9)
$$
$$
I_{7}^{np}\equiv \int_{0}^{\infty}y^{2}
{\left(y+\sqrt{x^{2}+y^{2}}\right)}^{-p-3}
{\Bigr(x^{2}+y^{2}\Bigr)}^{{p\over 2}-{n\over 2}-{3\over 2}} \; dy
\; \; \; \; ,
\eqno (C10)
$$
$$
I_{8}^{np}\equiv \int_{0}^{\infty}y
{\left(y+\sqrt{x^{2}+y^{2}}\right)}^{-p-3}
{\Bigr(x^{2}+y^{2}\Bigr)}^{{p\over 2}-{n\over 2}-{3\over 2}} \; dy
\; \; \; \; .
\eqno (C11)
$$
\vskip 100cm
\centerline {\bf REFERENCES}
\vskip 1cm
\item {${ }^{1}$}
K. Schleich, Phys. Rev. D {\bf 32}, 1889 (1985).
\item {${ }^{2}$}
J. Louko, Phys. Rev. D {\bf 38}, 478 (1988).
\item {${ }^{3}$}
J. Melmed, J. Phys. A {\bf 21}, L1131 (1988).
\item {${ }^{4}$}
I. G. Moss, Class. Quantum Grav. {\bf 6}, 759 (1989).
\item {${ }^{5}$}
G. V. M. Esposito, {\it Boundary Value Problems in Quantum cosmology}, Knight
Prize Essay, University of Cambridge, 1989 (unpublished).
\item {${ }^{6}$}
J. S. Dowker and I. G. Moss, Phys. Lett. {\bf 229 B}, 261 (1989).
\item {${ }^{7}$}
T. P. Branson and P. B. Gilkey, Commun. in Partial Differential Equations
{\bf 15}, 245 (1990).
\item {${ }^{8}$}
P. D. D'Eath and G. V. M. Esposito, {\it The Effect of Boundaries in One-Loop
Quantum Cosmology}, to appear in Proceedings of the Italian IX
National Conference of General Relativity and Gravitational Physics, Capri,
25-28 September 1990, DAMTP preprint R-90/20.
\item {${ }^{9}$}
I. G. Moss and S. Poletti, Nucl. Phys. {\bf B 341}, 155 (1990); Phys. Lett.
{\bf 245 B}, 355 (1990).
\item {${ }^{10}$}
A. O. Barvinsky and A. Yu. Kamenshchik, Class. Quantum Grav. {\bf 7}, L 181
(1990).
\item {${ }^{11}$}
S. Poletti, Phys. Lett. {\bf 249 B}, 249 (1990).
\item {${ }^{12}$}
H. C. Luckock
(Manchester University Preprint MC TH 90/18, 1990).
\item {${ }^{13}$}
M. F. Atiyah, {\it Collected Works}, Vol. 4 (Clarendon Press, Oxford, 1988)
359.
\item {${ }^{14}$}
P. D. D'Eath and J. J. Halliwell, Phys. Rev. D {\bf 35}, 1100 (1987).
\item {${ }^{15}$}
J. B. Hartle and S. W. Hawking, Phys. Rev. D {\bf 28}, 2960 (1983).
\item {${ }^{16}$}
S. W. Hawking, Nucl. Phys. {\bf B 239}, 257 (1984).
\item {${ }^{17}$}
P. D. D'Eath and G. V. M. Esposito, {\it Spectral Boundary Conditions in
One-Loop Quantum Cosmology}, paper in preparation.
\item {${ }^{18}$}
S. W. Hawking and C. N. Pope, Nucl. Phys. {\bf B 146}, 381 (1978).
\item {${ }^{19}$}
J. Kupsch and W. D. Thacker, Fortschr. Phys. {\bf 38}, 35 (1990).
\item {${ }^{20}$}
P. Breitenlohner and D. Z. Freedman, Ann. Phys. {\bf 144}, 249 (1982).
\item {${ }^{21}$}
S. W. Hawking, Phys. Lett. {\bf 126 B}, 175 (1983).
\item {${ }^{22}$}
R. Penrose and W. Rindler, {\it Spinors and Space-Time, Vol. 1 and 2}
(Cambridge University Press, New York, 1984-1986).
\item {${ }^{23}$}
G. V. M. Esposito, Ph. D. Thesis, University of Cambridge, 1990
(in preparation).
\item {${ }^{24}$}
H. C. Luckock and I. G. Moss, Class. Quantum Grav. {\bf 6}, 1993 (1989).
\item {${ }^{25}$}
P. D. D'Eath, in {\it Supersymmetry and its Applications}, edited by
G. W. Gibbons, S. W. Hawking and P. K. Townsend (Cambridge University Press,
New York, 1986).
\item {${ }^{26}$}
C. Itzykson and J. B. Zuber, {\it Quantum Field Theory} (McGraw-Hill Book Co,
New York, 1985).
\item {${ }^{27}$}
B. Allen, Nucl. Phys. {\bf B 226}, 228 (1983).
\item {${ }^{28}$}
N. M. J. Woodhouse, Class. Quantum Grav. {\bf 2}, 257 (1985).
\item {${ }^{29}$}
M. Reed and B. Simon, {\it Methods of Modern Mathematical Physics, Vol. 2}
(Academic Press, New York, 1975).
\item {${ }^{30}$}
S. W. Hawking, Commun. Math. Phys. {\bf 55}, 133 (1977).
\item {${ }^{31}$}
I. S. Gradshteyn and I. M. Ryzhik, {\it Table of Integrals, Series and
Products} (Academic Press, New York, 1965).
\item {${ }^{32}$}
M. Abramowitz and I. A. Stegun, {\it Handbook of Mathematical Functions}
(Dover, New York, 1964).
\item {${ }^{33}$}
M. V. Berry and R. J. Mondragon, Proc. R. Soc. Lond. A {\bf 412}, 53 (1987).
\item {${ }^{34}$}
L. V. Ahlfors, {\it Complex Analysis} (Mc Graw-Hill, New York, 1966).
\item {${ }^{35}$}
E. C. Titchmarsh, {\it The Theory of Functions}
(Oxford University Press, London, 1939).
\item {${ }^{36}$}
A. Ivi\'c, {\it The Riemann Zeta-Function} (John Wiley and Sons, New
York, 1985).
\item {${ }^{37}$}
F. W. J. Olver, Philos. Trans. Roy. Soc. London, Ser. {\bf A 247}, 328
(1954).
\item {${ }^{38}$}
R. Wong, {\it Asymptotic Approximations of Integrals} (Academic Press,
New York, 1989).
\item {${ }^{39}$}
J. D. Romano and R. S. Tate, Class Quantum Grav. {\bf 6}, 1487 (1989).
\item {${ }^{40}$}
K. Schleich, Class. Quantum Grav. {\bf 7}, 1529 (1990).
\item {${ }^{41}$}
M. Henneaux, Phys. Rep. {\bf 126}, 1 (1985).
\item {${ }^{42}$}
J. B. Hartle and K. Schleich, in {\it Quantum Field Theory and Quantum
Statistics : Essays in Honour of the Sixtieth Birthday of E. S. Fradkin,
Vol. 2}, edited by I. A. Batalin, G. A. Vilkovisky and C. J. Isham
(Adam Hilger, Bristol, 1987).
\bye